\documentclass[a4paper,fleqn,usenatbib]{mnras}
\usepackage{newtxtext,newtxmath}

\usepackage[T1]{fontenc}
\usepackage{ae,aecompl}
\usepackage{graphicx}	
\usepackage{amssymb}	
\usepackage{booktabs}	
\usepackage[percent]{overpic} 
\usepackage{ulem}
\usepackage{url}

\usepackage[dvipsnames]{xcolor}

\newcommand\mydots{\hbox to 1em{.\hss.\hss.}} 

\definecolor{meridithgreen}{RGB}{0, 150, 0}

\def\Dnu{$\Delta\nu$}
\def\LB{$\lambda$\,Boo}
\def\dsct{$\delta$\,Sct}
\def\teff{$T_{\rm eff}$}
\def\logL{$\log L$}

\usepackage{tablefootnote,footnotehyper} 
\usepackage{array}
\newcolumntype{H}{>{\setbox0=\hbox\bgroup}c<{\egroup}@{}} 

\title[Asteroseismology of HD\,139614]{A precise asteroseismic age and metallicity for HD\,139614: a pre-main-sequence star with a protoplanetary disc in Upper-Centaurus Lupus}

\author[Murphy et al.]{
Simon J. Murphy,$^{1,2}$\thanks{E-mail: simon.murphy@sydney.edu.au (SJM), Meridith.Joyce@anu.edu.au (MJ)}
Meridith Joyce,$^{3,4}$
Timothy R. Bedding$^{1,2}$, Timothy R. White$^{1,2}$ \and Mihkel Kama$^{5,6}$
\\
$^{1}$ Sydney Institute for Astronomy (SIfA), School of Physics, University of Sydney, NSW 2006, Australia\\
$^{2}$ Stellar Astrophysics Centre, Department of Physics and Astronomy, Aarhus University, 8000 Aarhus C, Denmark\\
$^{3}$ Research School of Astronomy and Astrophysics, Australian National University, Canberra, ACT 2611, Australia\\
$^{4}$ ARC Centre of Excellence for All Sky Astrophysics in 3 Dimensions (ASTRO 3D), Australia\\
$^{5}$ Department of Physics and Astronomy, University College London, Gower Street, London, WC1E 6BT, UK\\
$^{6}$ Tartu Observatory, University of Tartu, Observatooriumi 1, 61602 T\~{o}ravere, Tartumaa, Estonia
}

\date{Accepted XXX. Received YYY; in original form ZZZ}
\pubyear{2015}
\begin{document}
\label{firstpage}
\pagerange{\pageref{firstpage}--\pageref{lastpage}}
\maketitle

\begin{abstract}
HD\,139614 is known to be a $\sim$14\,Myr-old, possibly pre-main-sequence star in the Sco-Cen OB association in the Upper Centaurus-Lupus subgroup, with a slightly warped circumstellar disc containing ring structures hinting at one or more planets. The star's chemical abundance pattern is metal-deficient except for volatile elements, which places it in the $\lambda$\,Boo class and suggests it has recently accreted gas-rich but dust-poor material.
We identify seven dipole and four radial pulsation modes among its $\delta$\,Sct pulsations using the TESS light curve and an \'echelle diagram. Precision modelling with the MESA stellar evolution and GYRE stellar oscillation programs confirms it is on the pre-main sequence. Asteroseismic, grid-based modelling suggests an age of $10.75\pm0.77$\,Myr, a mass of $1.52\pm0.02$\,M$_{\odot}$, and a global metal abundance of $Z=0.0100\pm0.0010$.
This represents the first asteroseismic determination of the bulk metallicity of a $\lambda$\,Boo star.
The precise age and metallicity offer a benchmark for age estimates in Upper Centaurus--Lupus, and for understanding disc retention and planet formation around intermediate-mass stars.
\end{abstract}




\section{Introduction}
\label{sec:intro}

There are few initiatives in stellar astrophysics that would not benefit from better ages. Better ages for open clusters would improve benchmarks for almost every age determinant \citep{soderblom2010}, with impacts reaching as far as anchoring the initial-final mass relation of white dwarfs \citep{marigoetal2020}. Better ages from asteroseismology are reforming our understanding of angular momentum transport in stars \citep{aertsetal2019,eggenbergeretal2019,ouazzanietal2019,denhartoghetal2020}, and better ages for large populations of galactic field stars help to trace the chemical evolution of the Galaxy \citep{bland-hawthornetal2019,haydenetal2020,linetal2020}.
The formation and evolution of protoplanetary discs is another area where precise ages matter, with disc properties evolving on timescales of 3--5\,Myr \citep{haischetal2001,krausetal2012}.

The evolution of young stars, their discs, and their planets are intimately connected. For example, chemical peculiarities in \LB\ stars probably arise from a filtering of dust from gas in planet-forming discs \citep{kamaetal2015}. For unclear reasons, some such discs appear to survive exceptionally long, for $\gtrsim 5$ or even $10$\,Myr \citep[e.g.,][]{vioqueetal2018,flahertyetal2019}. Accurate measurements of the stellar mass, age, and intrinsic metallicity
are crucial constraints for the theoretical work connecting stellar mass build up, planet formation including migration, and disc dissipation \citep{gravitycollaboration2019}. Asteroseismology provides a unique and independent way of measuring those parameters. Here, we use a light curve from TESS \citep{rickeretal2015} in an asteroseismic analysis of the young, disc-hosting, chemically peculiar star HD\,139614

HD\,139614 is located in Upper Centaurus--Lupus (UCL), a subdivision of the Sco--Cen association which, at 140\,pc, is the nearest region to the Sun with high-mass star formation \citep{blaauw1946,dezeeuwetal1999}. The relative youth of UCL (median age = $16\pm2$\,Myr, \citealt{pecaut&mamajek2016}), its observed 1$\sigma$ age spread of $\sim$7\,Myr \citep{mamajeketal2002,preibisch&mamajek2008,pecaut&mamajek2016}, and its proximity make it an important laboratory for high-angular-resolution studies of planet-forming discs of various ages \citep[e.g.,][]{ansdelletal2016,liemansifryetal2016,barenfeldetal2017}, each of which provide a snapshot of disc evolution. There is mounting evidence that quite massive dust discs can survive for a wide range of ages \citep[e.g.,][]{pfalzneretal2014}, but determining those ages is difficult. Ages of stars within UCL are known to vary with both location within the association and with stellar mass \citep{hillenbrandetal2008,pecaut&mamajek2016}, which means there is a need for age calibrators throughout the region.

HD\,139614 is a well-studied member of UCL (membership probability 99.9\%, \citealt{gagneetal2018}) whose protoplanetary disc is known to have dust gaps and a cavity \citep{meeusetal1998,matteretal2014,matteretal2016,carmonaetal2017}. Scattered-light imagery shows a warped disc \citep{muro-arenaetal2020} that may be evidence of embedded planets on inclined orbits \citep{nealonetal2018,nealonetal2019}. Dust traps next to empty gaps or an inner cavity, which are also seen in the HD\,139614 system, may allow large quantities of dust to survive for much longer than in many other disks \citep{pinillaetal2018, pinillaetal2020}. Embedded planets are able to trap dust in the disc and prevent its accretion onto the star \citep{jermyn&kama2018}, leaving fingerprints in the stellar spectrum of dust-poor accretion \citep{kamaetal2015}. This is one hypothesized mechanism for producing the chemical peculiarities that typify the \LB\ class \citep{venn&lambert1990,watersetal1992,king1994}, which exhibit a depletion of up to 2\,dex in refractory elements alongside solar abundances of volatiles \citep{baschek&slettebak1988,andrievskyetal2002,kampetal2001}. Indeed, VLTI and ALMA observations confirm the existence of planets embedded in the discs of some $\lambda$\,Boo stars \citep[e.g.][]{matteretal2016,fedeleetal2017,cugnoetal2019,tocietal2020}. Although no planets are yet confirmed for HD\,139614, it is known to be a \LB\ star based on its photospheric abundances \citep{folsometal2012,murphyetal2015b}.

No conclusive assessment has yet been made as to whether \LB\ stars are
globally metal-poor objects, though strong statistical arguments have been made to the contrary \citep{paunzenetal2015,murphyetal2020b}. While spectroscopy is able to determine the photospheric metallicity, asteroseismology probes the stellar interior and can deliver precise bulk metallicities (e.g. \citealt{vansaders&pinsonneault2012}), because the latter affects the stellar structure to which the oscillation frequencies are sensitive (e.g. \citealt{guziketal1998,aertsetal2010}). The application of asteroseismic techniques to \LB\ stars is therefore highly sought after \citep{paunzenetal1997b}. In addition, \LB\ stars may make particularly good asteroseismic targets since they are twice as likely as normal stars of the same effective temperature to pulsate as $\delta$\,Sct stars \citep{murphyetal2020b}, which are intermediate-mass stars pulsating predominantly in pressure modes \citep{breger2000}. While many studies have been dedicated to the detection of pulsations in \LB\ stars \citep[e.g.][]{paunzenetal1998b,koenetal2003,bregeretal2006,paunzenetal2015}, somewhat fewer have succeeded in the first asteroseismic step of identifying the pulsation modes \citep{paunzenetal2002a,mkrtichianetal2007,beddingetal2020,murphyetal2020b}, and only \citet{casasetal2009} have made any inferences on the stellar interior.

For a long time, the inability to successfully identify the radial order ($n$) and the spherical degree ($\ell$) of the pulsations of \dsct\ stars was a major barrier to progress \citep[e.g.][]{bregeretal2004}; many oscillation modes were observed, but few had reliable mode identifications. The situation grew worse with space photometry of \dsct\ stars from CoRoT \citep{garciahernandezetal2009,porettietal2010,mantegazzaetal2012} and \textit{Kepler} \citep[e.g.][]{uytterhoevenetal2011}, where unprecedented photometric precision revealed many more oscillation frequencies. However, mode identifications are traditionally obtained via time-series spectroscopy \citep{balona1986} or multi-colour photometry \citep{watson1988} and techniques are limited with single pass-band photometry alone \citep[cf.][]{petersen&christensen-dalsgaard1996}, but a recent breakthrough made possible by the large number of stars observed at rapid cadence by TESS \citep{rickeretal2015} has enabled mode identification for some high-frequency \dsct\ stars \citep{beddingetal2020}.

At high radial orders, pressure modes (p\:modes) are asymptotically spaced in frequency \citep{unnoetal1989} by the so-called large spacing, \Dnu. Most \dsct\ stars oscillate at low frequencies, corresponding to low radial orders and periods longer than $\sim$1\,hr \citep{rodriguezetal2000,balona&dziembowski2011}, where spacings are not regular and modes are therefore difficult to identify. \citet{beddingetal2020} identified $\sim$60 TESS \dsct\ stars oscillating at high frequencies with regular frequency patterns. Gaia DR2 parallaxes \citep{gaiacollaboration2018a} showed that these stars were close to the ZAMS, and \citet{beddingetal2020} found that \LB\ stars were overrepresented in this sample, consistent with the idea that many \LB\ stars are young objects.\footnote{Though not all are young; \citet{murphy&paunzen2017} showed with DR1 parallaxes that \LB\ stars are found at all main-sequence ages, and they may even exhibit infrared excesses at a wide range of main-sequence ages \citep{grayetal2017,murphyetal2020c}.} A complete examination of southern \LB\ stars observed in the first year of TESS observations confirmed regular frequency spacings in many of their Fourier spectra, including in HD\,139614, and the fundamental mode was identifiable in half the sample \citep{murphyetal2020b}.

The recent progress on \dsct\ stars offers the exciting prospect of precise asteroseismic ages for intermediate mass stars and any discs or planets they may host. Until now, such ages have generally been limited to lower-mass stars via, e.g., empirical relations with stellar spindown \citep{barnes2007}, activity decay \citep{baliunasetal1995}, lithium depletion \citep{skumanich1972}, or isochrone fitting (\citealt{valenti&fischer2005}; see \citealt{soderblom2010} for a review), whereas for \dsct\ stars, these methods are not generally applicable \citep{russell1995,pedersenetal2017,vansadersetal2019,zwintz2020}. A related class of stars, the $\gamma$\,Doradus stars that pulsate in gravity modes \citep{guziketal2000b,grigahceneetal2010a}, has recently shown promise for asteroseismic ages because the gravity modes are very sensitive to chemical discontinuities in the near-core region \citep{miglioetal2008,aerts2015,vanreethetal2015b}. Unfortunately, the $\gamma$\,Dor instability strip is narrow \citep{dupretetal2005b} and such detailed asteroseismic information is only extractable from about 10\% of $\gamma$\,Dor pulsators \citep{glietal2020a}. With the exception of certain \dsct{}--$\gamma$\,Dor hybrids \citep[e.g.][]{kurtzetal2014,saioetal2015}, the best \dsct\ ages have often come from pre-main-sequence (pre-MS) members of the class \citep{zwintzetal2014b}. While $\sim$50 pre-MS \dsct\ stars are known \citep{zwintz2017}, a lack of reliable mode identifications has meant that determining their evolutionary status has relied upon classical parameters such as effective temperature (\teff), luminosity ($L$), and metallicity ($Z$), and hindered the use of the pulsations to infer metallicities or to improve the stellar models \citep{ripepietal2011,ripepietal2015,chen&li2018,chen&li2019}.

In this paper, we perform mode identification and determine an asteroseismic age and metallicity for HD\,139614, which we conclude is a pre-MS star near to the zero-age main sequence (ZAMS). In Sec.\,\ref{sec:lit} we describe and summarise the stellar parameters of HD\,139614 available in the literature. We describe our analysis of the TESS photometry, including mode identification, in Sec.\,\ref{sec:puls}. We describe our modelling method in Sec.\,\ref{sec:models}, and our results in Sec.\,\ref{sec:results}. Detailed discussions are presented in Sec.\,\ref{sec:discussion} before the conclusions in Sec.\,\ref{sec:conclusions}.


\section{HD\,139614 in the literature}
\label{sec:lit}

The discovery of emission lines and a disc around HD\,139614 has led to it being very well studied \citep{henize1976,meeusetal1998}, with several studies dedicated to observations and modelling of its disc \citep{matteretal2014,menuetal2015,matteretal2016,carmonaetal2017,seok&li2017,lawsetal2020,muro-arenaetal2020}. In this section, we describe the stellar atmospheric parameters available from the literature for HD\,139614, which we summarise in Table\:\ref{tab:puls}.

Although its abundances were determined by \citet{dunkinetal1997} and again by \citet{acke&waelkens2004}, the first suggestion that HD\,139614 is a \LB\ star came from a spectroscopic study of some Herbig Ae stars by \citet{folsometal2012}. Its abundances were found to be typical of \LB\ stars and it was accepted as a member of the class shortly thereafter \citep{murphyetal2015b}. With respect to the \citet{asplundetal2009} solar photosphere, the \citet{folsometal2012} Fe abundance for HD\,139614 is [Fe/H] = $-0.57\pm0.13$; other abundances relevant to this classification are [C/H] = $-0.18$, [N/H] =  $0.00$, and [O/H] = $-0.02$. This metallicity ([Fe/H]) has generally been used for HD\,139614 in studies of its disc or for computing isochrones to infer the stellar age, but it is likely that this metallicity applies neither to the whole disc \citep{kamaetal2015} nor the bulk of the star \citep{paunzenetal2015,murphyetal2020b}, because it is the manifestation of a surface peculiarity originating in a dust cavity in the disc. In this work, we did not assume any particular metallicity for the star. Instead, we explored a range of metallicities around the solar value $Z=0.014$, which appears to be appropriate for stars in UCL \citep{lastennet&valls-gabaud2002,preibisch&mamajek2008}. Further details are given in Sec.\,\ref{sec:models}.

Spectroscopic effective temperatures of 7600--8250\,K have been determined for HD\,139614 by various groups \citep{dunkinetal1997,acke&waelkens2004,folsometal2012,fairlambetal2015}, with the most recent values being \teff{} $=7600\pm300$ and $7750\pm250$\,K, respectively. A recent \teff{} from Str\"omgren photometry is in good agreement with this \citep[$7626\pm153$\,K;][]{murphyetal2020b}. We adopt $T_{\rm eff} = 7650\pm200$\,K for this study.

A precise parallax of $7.4243\pm0.0533$\,mas is available from Gaia DR2 \citep{gaiacollaboration2018a}, for use in calculating a stellar luminosity. \citet{vioqueetal2018} determined $\log L/L_{\odot}$ = $0.773^{+0.032}_{-0.010}$, which equates to $L = 5.93^{+0.45}_{-0.14}$\,L$_{\odot}$. Conversely, \citet{murphyetal2020b} determined $L = 6.78\pm0.29$\,L$_{\odot}$ (\logL{} $=0.83$), which is consistent with the value of \citet[][\logL{} $=0.82$]{fairlambetal2015}. The exact cause for the difference is unknown, but could arise from different assumptions on metallicity or extinction in the luminosity calculation. In this work we adopt $L$ = 6.7$\pm$0.5\,L$_{\odot}$ ($\log L/{\rm L}_{\odot} = 0.83\pm0.03$); exploration of a broader luminosity range ($>1\sigma$) using seismic constraints shows that the lower luminosity of \citet{vioqueetal2018} yields a poor fit (Sec.\,\ref{sec:results}). It is also noteworthy that
\citet{fairlambetal2015} found a negligible accretion luminosity of 
$-0.10^{+0.20}_{-0.31}$\,L$_{\odot}$.

The stellar age has been determined by many authors by reference to isochrones, with varying degrees of precision \citep[e.g.][]{folsometal2012,fairlambetal2015,gagneetal2018,vioqueetal2018}. The median age, which also has the smallest uncertainty, is $14.5^{+1.4}_{-3.6}$\,Myr \citep{vioqueetal2018}. This is consistent with the age map of UCL by \citet[][their figure~9]{pecaut&mamajek2016}, which suggests an age of $14\pm2$\,Myr for HD\,139614. In our seismic modelling (Sec.\,\ref{sec:models}), we applied an age prior of $< 30$\,Myr.

HD\,139614 is an unusually slow rotator for a star of its mass ($1.49\pm0.07$\,M$_{\odot}$; \citealt{fairlambetal2015}), with \mbox{$v\sin i = 24$--26\,km\,s$^{-1}$;} \citep{dunkinetal1997,acke&waelkens2004,folsometal2012,alecianetal2013a}. The rotational velocity distribution for A stars with mass 1.6\,M$_{\odot}$ \citep{zorec&royer2012}, has a mode at $v_{\rm eq} = 150$\,km\,s$^{-1}$ (their figure 7), with the 1$\sigma$ and 2$\sigma$ ranges being $\sim$110--225, and $\sim$90--300\,km\,s$^{-1}$, respectively. This suggests that HD\,139614 is either an intrinsically slow rotator (i.e. $v_{\rm eq}$ is small) or it has a low inclination, $i$, or both. Disc modelling shows that the disc inclination is low, at $20\pm2$\,deg \citep{matteretal2014}, although there are misalignments within the disc of a few degrees \citep{muro-arenaetal2020}. The stellar inclination and the disc inclination are not necessarily equal \citep{ansdelletal2020}, particularly if there is a companion orbiting within the disc \citep{matsakos&koenigl2017}. But if the star and disc are aligned, then HD\,139614 has $v_{\rm eq} \sim 70$\,km\,s$^{-1}$, which is still slow for its mass. Since this is a small fraction of the Keplerian break-up velocity ($\Omega = 0.16 \Omega_{\rm K}$), we expect the star to have negligible oblateness.

\begin{table}
\centering
\caption{A summary of the stellar parameters for HD\,139614 inferred from the literature. Some of these are used in our modelling (see Sec.\,\ref{sec:models} for details).}
\label{tab:params}
\begin{tabular}{l c c c}
\toprule
Parameter & Value & Units & Section \\
\midrule
$T_{\rm eff}$ & $7650\pm200$ & K & Sec.\,\ref{sec:lit}\\
$L$ & $6.7\pm0.5$ & ${\rm L}_{\odot}$ & Sec.\,\ref{sec:lit}\\
$Z$ & $\sim$0.014 & --- & Sec.\,\ref{sec:lit}\\
age & $<$30 & Myr & Sec.\,\ref{sec:lit}\\
$v \sin i$ & 24--26 & km\,s$^{-1}$ & Sec.\,\ref{sec:lit}\\
$v_{\rm eq}$ & $70\pm10$ & km\,s$^{-1}$ & Sec.\,\ref{sec:lit}\\
$\Delta\nu$ & $6.83\pm0.02$ & d$^{-1}$ & Sec.\,\ref{sec:puls}\\
\bottomrule
\end{tabular}
\end{table}

\section{Pulsation properties from TESS photometry}
\label{sec:puls}

TESS observed HD\,139614 for 25.2\,d at 2-min cadence in Sector 12 (2019 May/Jun) during the first year of its nominal mission (it will observe it for another $\sim$27\,d in the extended mission in Sector 38 (2021 Apr/May). TESS light curves are available in Simple Aperture Photometry (SAP) and Pre-Search Data Conditioning (PDC) versions; the latter has undergone processing to remove instrumental signals \citep{jenkinsetal2016}. We downloaded both versions from the MAST portal\footnote{\url{https://mast.stsci.edu/}} using the {\sc lightkurve} package \citep{lightkurvecollaboration2018}, and after visually checking that the PDC algorithm had not removed astrophysical signals, we analysed this version of the light curve. To minimise the uncertainties, we subtracted the mid-time of the observations, 2\,458\,640.273\,d.

\begin{figure*}
\begin{center}
\includegraphics[width=0.98\textwidth]{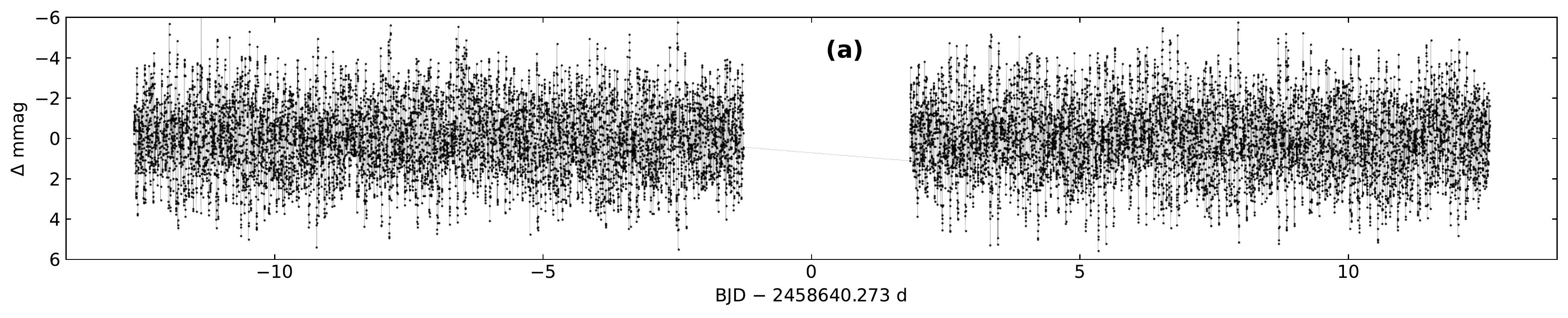}
\includegraphics[width=0.98\textwidth]{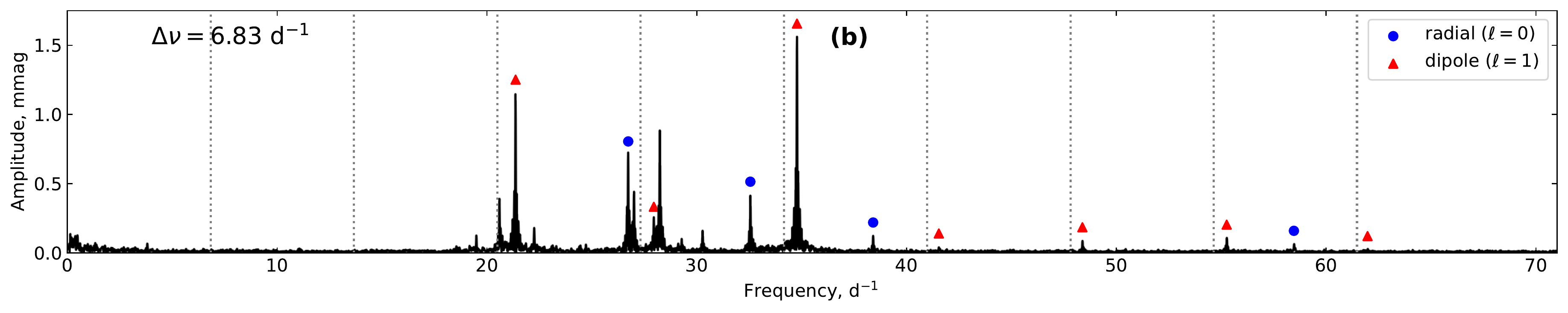}
\caption{{\bf (a):} TESS Sector 12 light curve of HD\,139614, showing \dsct\ pulsations with 0.4--1.2\,h periods. {\bf (b):} The Fourier transform of the TESS light curve. The frequencies show a characteristic spacing, especially at high frequency, which we identify as the large separation of \Dnu\ $=6.83$\,d$^{-1}$. Mode IDs are also shown for the modes we could identify.}
\label{fig:puls}
\end{center}
\end{figure*}

The light curve shows peak-to-peak variability of $\pm$5\,mmag (Fig.\,\ref{fig:puls}a), with periods in the range 0.4--1.2\,h typical of \dsct\ stars near the ZAMS. There is no evidence of transient dimming by circumstellar material, as seen in UXOR variables \citep{herbstetal1994} and their less extreme counterparts, the `young dippers' \citep{codyetal2014,ansdelletal2016,staufferetal2017}. This supports the view that the disc is seen nearly face-on \citep{meeusetal1998}.

\begin{figure}
\begin{center}
\includegraphics[width=0.48\textwidth]{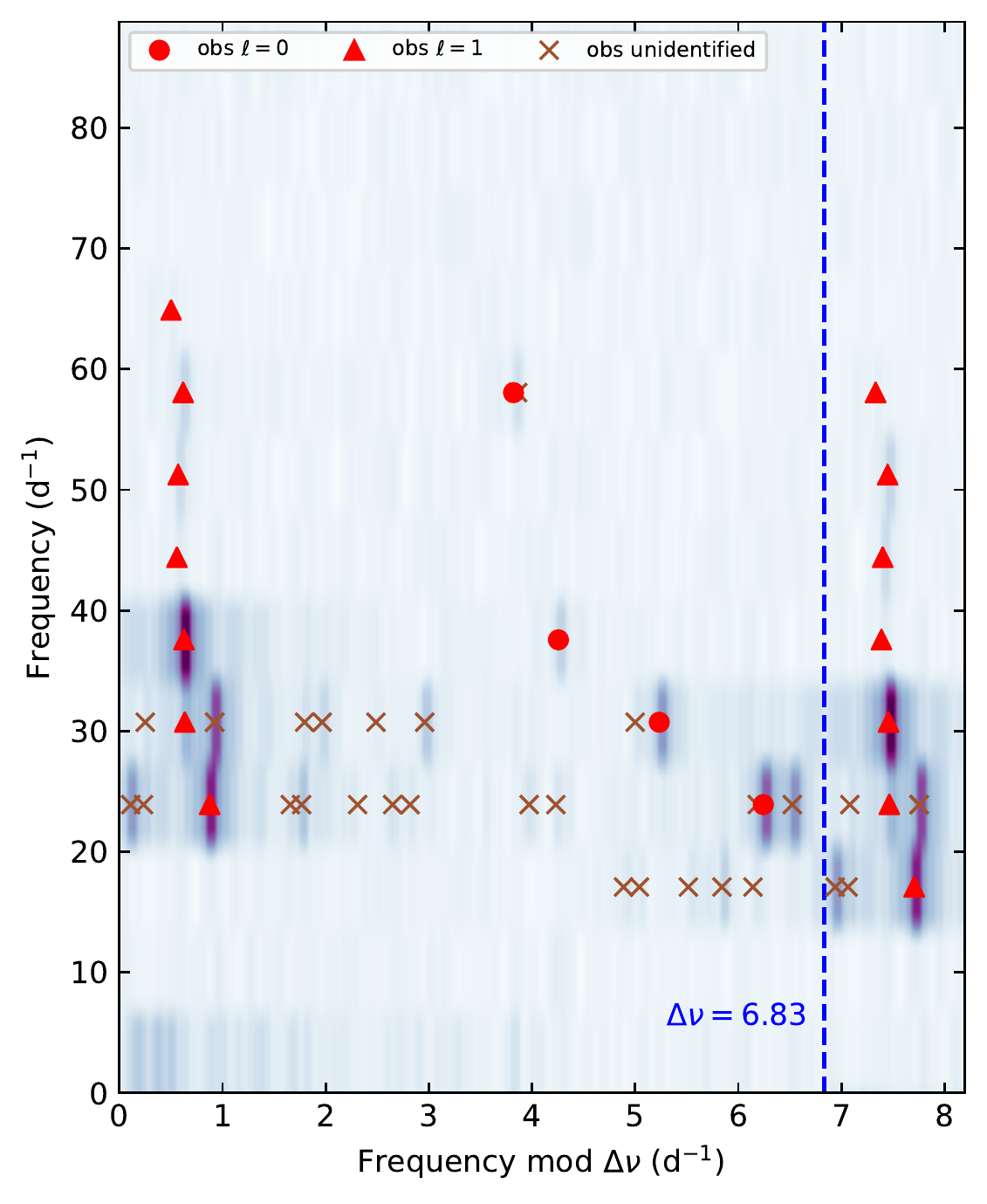}
\caption{\'Echelle diagram of HD\,139614, with large separation, \mbox{$\Delta\nu=6.83$\,d$^{-1}$}. A repeated overlap region is shown on the right for clarity. The greyscale indicates the Fourier amplitude at that frequency, and some smoothing has been applied. Circles show radial modes, triangles show dipole modes, and observed peaks without corresponding mode IDs are shown as crosses.}
\label{fig:echelle}
\end{center}
\end{figure}

\citet{beddingetal2020} showed how the frequencies of some \dsct\ stars can lie on ridges in an \'echelle diagram when the correct large spacing (\Dnu{}) is used, according to the angular degree of the modes. \citet{murphyetal2020b} applied this technique to HD\,139614, showing two clear ridges at \Dnu\ $= 6.83$\,d$^{-1}$. The \'echelle diagram is shown in Fig.\,\ref{fig:echelle}. Using only these observations, we confirm the large spacing of $\Delta\nu=6.83\pm0.02$\,d$^{-1}$ ($79.05\pm0.23$\,$\upmu$Hz), which is determined as the value that makes the dipole mode ridge vertical over orders $n = 3$--8 \citep{beddingetal2020}, with convention that $n=1$ is the fundamental mode. Since $n=8$ is below the canonical ``asymptotic regime'' for p\:modes, our observed $\Delta\nu$ can be expected to differ from the typical output of stellar evolution codes such as MESA (discussed in detail in Sec.\,\ref{ssec:seismic_models}). \citet{beddingetal2020} found that the radial mode ridge curves at low frequency, while the dipole ridge remains vertical, providing a way to identify these ridges. In this way, we identified the dipole modes as the peaks along the left side of Fig.\,\ref{fig:echelle}, and the radial modes as the ones down the centre and curving to the right. Some later refinement to this mode identification was made by comparing with preliminary models (Sec.\,\ref{sec:models}); our final mode IDs are shown in Figs\,\ref{fig:puls}b and \,\ref{fig:echelle}.

The $\ell=1$ modes appear to be singlets, rather than rotationally split doublets or triplets. As mentioned in Sec.\,\ref{sec:lit}, the inclination of the star is expected to be low, so singlets corresponding to $m=0$ are expected for the dipole modes \citep{gizon&solanki2003}. This offers a loose constraint of $i<30^{\circ}$, independent of the $i_{\rm disc}=20$\,deg measurement, and is consistent with star--disc alignment.


For precise frequency modelling, we determined the pulsation frequencies and amplitudes via simultaneous least-squares fitting using {\small PERIOD04} \citep{lenz&breger2004}. We extracted peaks above 5\,d$^{-1}$ down to an amplitude limit of 0.03\,mmag, corresponding to signal-to-noise ratios (SNR) of 3--4 for the weakest peaks. At frequencies below 5\,d$^{-1}$, there are few significant peaks (Fig.\,\ref{fig:puls}) and no means of identifying the modes or distinguishing them from instrumental signals. The extracted frequencies are shown in Table\:\ref{tab:puls} along with their identifications. The table also includes a weak peak at 61.97\,d$^{-1}$, which could be identified because of its location on the dipole mode ridge in the \'echelle diagram.

We determined uncertainties using a 100-process Monte Carlo simulation in {\small PERIOD04}.
The Fourier noise (0.0068\,mmag) was taken to be the median Fourier amplitude in the residuals of our multi-frequency fit at frequencies > 5\,d$^{-1}$, and was used to calculate the SNR of each peak. The observational frequency uncertainties are as low as $10^{-4}$\,d$^{-1}$ in some cases, with strong dependence on pulsation amplitude, as expected \citep{montgomery&odonoghue1999,kjeldsen&bedding2012}.



\begin{table}
\centering
\caption{Significant frequencies in the TESS light-curve of HD\,139614. The first column is a list of labels used in the text. Modes lying on the radial or dipole ridge have been identified with their radial order $n$ and degree $\ell$. The fundamental radial mode, with $n=1$ and $\ell=0$, is not observed. Entries are ordered by mode ID, then by those mentioned in the text (F12--F14), then by frequency.}
\label{tab:puls}
\begin{tabular}{c r r r r r c c}
\toprule
Label & \multicolumn{1}{c}{Freq.} & \multicolumn{1}{c}{$\sigma$Freq.} & Amp. & $\sigma$Amp. & SNR & $n$ & $\ell$ \\
 & d$^{-1}$ & d$^{-1}$ & mmag & mmag & & & \\
 \midrule
F1 & 21.36513 & 0.00013 & 1.152 & 0.006 & 169.4 & 1 & 1 \\
F2 & 27.95328 & 0.00041 & 0.256 & 0.005 & 37.6 & 2 & 1 \\
F3 & 34.77609 & 0.00006 & 1.557 & 0.005 & 228.9 & 3 & 1 \\
F4 & 41.53747 & 0.00274 & 0.040 & 0.007 & 5.9 & 4 & 1 \\
F5 & 48.37933 & 0.00161 & 0.085 & 0.006 & 12.5 & 5 & 1 \\
F6 & 55.25743 & 0.00132 & 0.104 & 0.006 & 15.3 & 6 & 1 \\
F7 & 61.97086 & 0.15366 & 0.021 & 0.007 & 3.1 & 7 & 1 \\
F8 & 26.73244 & 0.00020 & 0.705 & 0.027 & 103.6 & 2 & 0 \\
F9 & 32.55347 & 0.06198 & 0.414 & 0.136 & 60.9 & 3 & 0 \\
F10 & 38.40601 & 0.00100 & 0.119 & 0.006 & 17.5 & 4 & 0 \\
F11 & 58.46088 & 0.00321 & 0.059 & 0.006 & 8.6 & 7 & 0 \\
F12 & 28.24243 & 0.00017 & 0.885 & 0.006 & 130.1 & & \\
F13 & 27.01307 & 0.00027 & 0.458 & 0.007 & 67.3 & & \\
F14 & 20.59936 & 0.00036 & 0.409 & 0.006 & 60.2 & & \\
F15 & 18.54655 & 0.00272 & 0.047 & 0.006 & 6.9 & & \\
F16 & 18.69868 & 0.05470 & 0.034 & 0.012 & 4.9 & & \\
F17 & 19.17648 & 0.00204 & 0.046 & 0.007 & 6.8 & & \\
F18 & 19.50079 & 0.00092 & 0.107 & 0.006 & 15.7 & & \\
F19 & 19.80056 & 0.00440 & 0.031 & 0.005 & 4.6 & & \\
F20 & 20.72404 & 0.00280 & 0.043 & 0.006 & 6.3 & & \\
F21 & 22.15081 & 0.00148 & 0.080 & 0.007 & 11.8 & & \\
F22 & 22.25558 & 0.00080 & 0.167 & 0.007 & 24.6 & & \\
F23 & 22.79990 & 0.13974 & 0.030 & 0.008 & 4.5 & & \\
F24 & 23.13894 & 0.00173 & 0.066 & 0.006 & 9.7 & & \\
F25 & 23.30843 & 0.03089 & 0.046 & 0.012 & 6.8 & & \\
F26 & 24.46210 & 0.31248 & 0.052 & 0.019 & 7.7 & & \\
F27 & 24.72077 & 0.96335 & 0.062 & 0.038 & 9.1 & & \\
F28 & 26.67124 & 0.57532 & 0.064 & 0.034 & 9.4 & & \\
F29 & 27.57177 & 0.00334 & 0.046 & 0.006 & 6.8 & & \\
F30 & 29.12031 & 0.00194 & 0.058 & 0.006 & 8.5 & & \\
F31 & 29.28361 & 0.00120 & 0.108 & 0.006 & 15.9 & & \\
F32 & 29.80605 & 0.01082 & 0.024 & 0.007 & 3.5 & & \\
F33 & 30.27912 & 0.00082 & 0.152 & 0.006 & 22.4 & & \\
F34 & 32.31897 & 1.26274 & 0.049 & 0.020 & 7.2 & & \\
F35 & 58.50036 & 0.00638 & 0.031 & 0.005 & 4.5 & & \\
\bottomrule
\end{tabular}
\end{table}

\section{Modelling Method}
\label{sec:models}
We performed grid-based modelling using stellar tracks computed with the Modules for Experiments in Stellar Astrophysics (MESA) software version 11701 and theoretical frequency spectra computed with the GYRE stellar oscillation program version 5.2 \citep{paxtonetal2011, paxtonetal2013, paxtonetal2015, paxtonetal2018, paxtonetal2019, townsend&teitler2013}. We computed the stellar tracks and oscillation spectra separately, using structural models from particular evolutionary timesteps as initial conditions for higher-precision oscillation calculations. The details of these two types of calculation are described below.

\subsection{Evolutionary Tracks}
\label{ssec:tracks}
Our grid considered variations over a number of parameters that dictate the star's evolutionary trajectory. Mass, global metal abundance ($Z$), and the convective mixing length ($\alpha_{\text{MLT}}$), were varied at initial resolutions of $0.1 M_{\odot}$, $0.001$, and 0.1\,$H_{\rm p}$,\footnote{pressure scale heights} respectively. Our best models are reported with final precisions of $0.01 M_{\odot}$ and $0.0005$ in $Z$.
We aimed first to isolate the parameter ranges and combinations that produced stellar tracks consistent with the luminosity and effective temperature of HD\,139614 (i.e. ``classical'' constraints) independently of assumptions about the star's age or seismic features.
Following the approach of \citet{joyce&chaboyer2018b}, grid resolution was increased adaptively in regions of the parameter space where good agreement was found with HD\,139614's classical observational constraints.

Our models adopted a simple photosphere (evaluated at an optical depth of $\tau = 2/3$) for atmospheric surface boundary conditions, and we tested the effects of different model atmosphere choices using the best fitting model (Sec.\,\ref{ssec:global_fits}). We used the scaled solar abundances and opacities of \citet{asplundetal2009}. We used the MESA equation of state (EOS), a basic nuclear reaction network with additional rates for $^{56}$Fe and $^{58}$Ni,\footnote{\texttt{basic\_plus\_fe56\_ni58.net, in MESA}} and the Henyey implementation of the mixing length formalism \citep{henyeyetal1965,paxtonetal2011}.
All of our stellar tracks assumed the solar value for the initial helium abundance, $Y=0.29$. Although there is a known degeneracy between mass, composition, and mixing length when attempting to reproduce the surface features of stars (cf. \citealt{joyce&chaboyer2018a}'s analysis of HD\,140283), our models were found to be insufficiently sensitive to metallicity to suggest that variations in $Y$ would have a noticeable impact on our results. Hence, we did not consider variations in $Y$. Our models were likewise found to be insensitive to the choice of mixing length at the 20\% level, making such considerations less important than in the cases studied by \citet{joyce&chaboyer2018a} and \citet{joyce&chaboyer2018b}.

\begin{figure}
    \centering
    \includegraphics[width=0.47\textwidth]{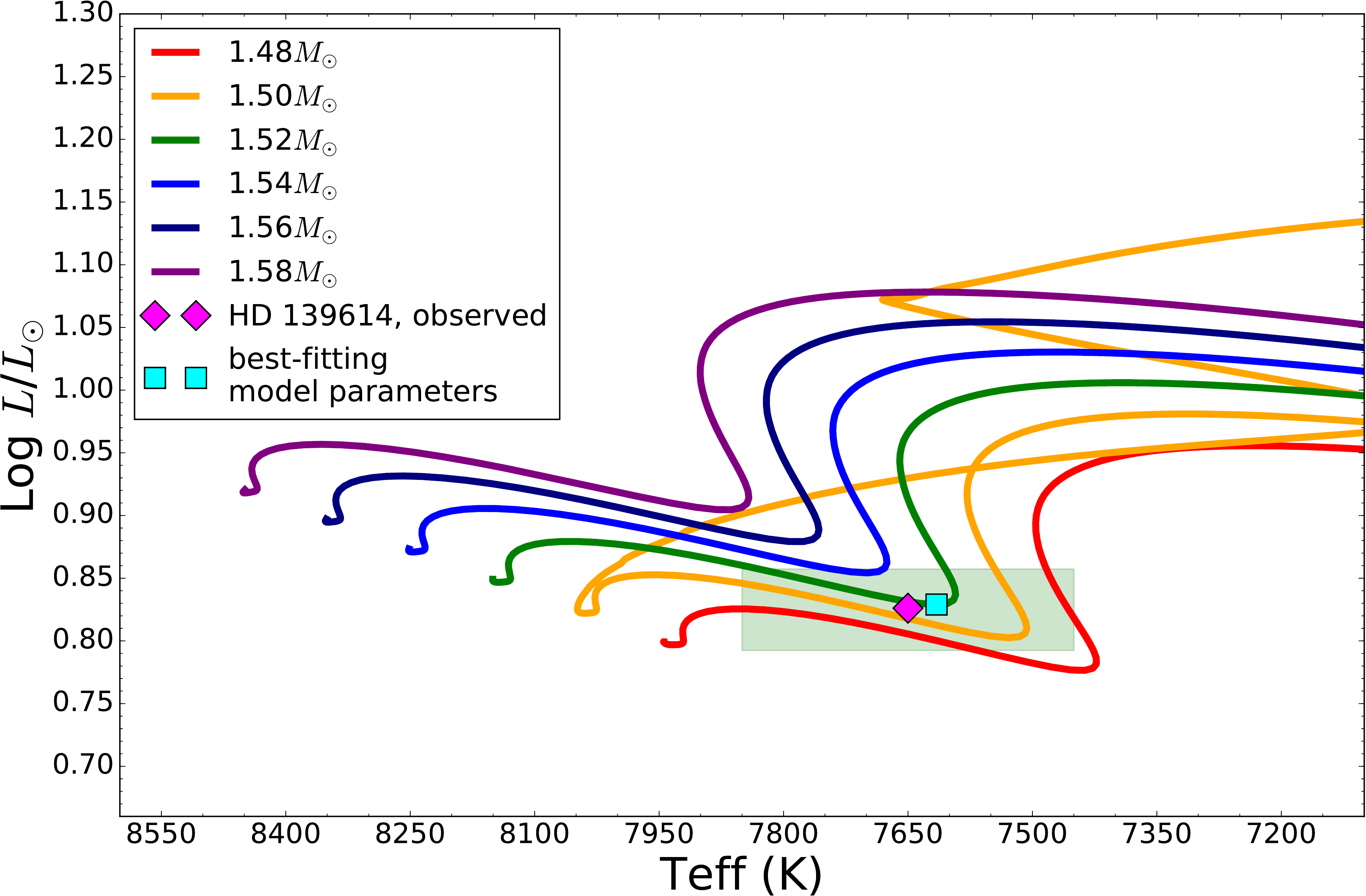}
    \caption{Select tracks with varying mass are shown in the HR diagram. The $1\sigma$ observational uncertainties are shown in the green shaded region. All tracks were terminated at an age of $30$\,Myr, with the exception of the $1.50$-M$_{\odot}$ model, shown in orange, to illustrate the main-sequence evolution.}
    \label{fig:hr_prems}
\end{figure}

The $1\sigma$ ranges of temperature and luminosity measurements (Table\:\ref{tab:params}) are shown on the HR diagram in Fig.\,\ref{fig:hr_prems}, along with five stellar tracks of different mass. All models shown have $\alpha_{\text{MLT}}=1.7$ and $Z=0.010$. All were terminated at an age of $30$\,Myr, with the exception of the track with mass 1.50\,$ M_{\odot}$, which was continued until an age of 1\,Gyr. As the 1.50-M$_{\odot}$ track demonstrates, models in this mass and metallicity regime are only consistent with the observational constraints during the pre-main sequence. However, it is straightforward to match the classical observables with models of different but plausible mass and metallicity at each of the three subsequent evolutionary phases: main sequence, convective turn off, and base of the subgiant branch. Without imposing age restrictions or taking seismic quantities into account, the range of possible masses, metallicities, and ages is quite large.
When seismology (in the form of MESA's $\Delta \nu$ approximation; see subsequent discussion) was even loosely considered, however, it became immediately clear that models in evolutionary stages beyond the pre-main sequence would not fit HD\,139614 unless they were infeasibly old---at least $0.8$\,Gyr. Given the additional context of literature estimates of the association's age (10--20\,Myr, Sec.\,\ref{sec:intro}; \citealt{pecaut&mamajek2016,luhman&esplin2020}), we imposed a generous $30$\,Myr age maximum on the tracks and restricted further analysis to pre-main sequence models only.

Stellar ages prior to the ignition of hydrogen burning are measured from a zero point defined by the choice of central temperature for the pre-main sequence model. For our case, the default setting of \texttt{pre\_ms\_T\_c} = 3d5 (i.e. $3\times10^5$\,K) was used. This choice is arbitrary but is widely adopted among MESA users and does not have a significant  effect on the measured age (see discussion in Sec.\,\ref{ssec:age_systematics}).

%

\begin{figure}
\centering
\begin{overpic}[width=0.47\textwidth]{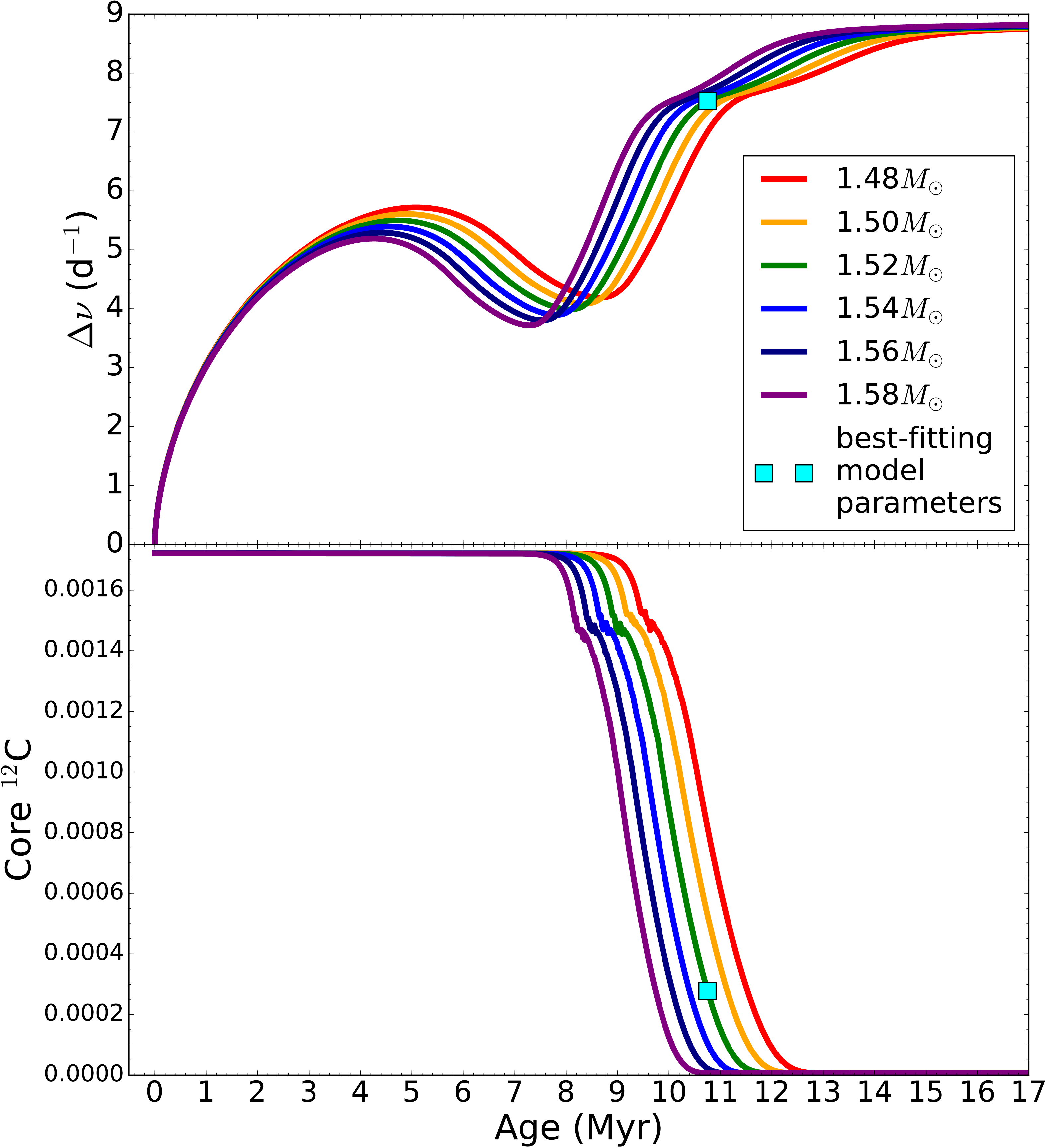}
\put (15,93) {\large {\bf (a)}}
\put (15,45) {\large {\bf (b)}}
\end{overpic}
\caption{{\bf (a):} The evolution of $\Delta \nu$ over time is shown along the pre-main sequence for six evolutionary tracks spanning a small range of masses. The point corresponding to our best-fitting model is indicated via marker. We note that the values of $\Delta \nu$ shown here are asymptotic values taken from MESA directly and are therefore approximate, computed from the stellar structure equations rather than from the oscillation equations. {\bf (b):} Evolution of the core carbon-12 content.}
\label{fig:dnu_age}
\end{figure}

Fig.\,\ref{fig:dnu_age}a demonstrates the rapid change in $\Delta \nu$ in the model's first $\sim$15\,Myr of evolution (see also \citealt{suranetal2001}). During this time, the star undergoes catalytic hydrogen fusion by proton capture with C$^{12}_{6}$, but completion of the CNO cycle does not yet occur, and C$^{12}_{6}$ is exhausted over a $\sim$2-Myr period (Fig.\,\ref{fig:dnu_age}b). The tracks in Fig.\,\ref{fig:dnu_age} have the same parameters as those in Fig.\,\ref{fig:hr_prems}, but demonstrate sensitivity to mass for only $\sim$12 of the first 15\,Myr. The sensitivity to $\Delta \nu$, however, is present regardless of other parameters; likewise, our full seismic models are also sensitive to the rapidly evolving structure during this period. It is this sensitivity that allowed us to date our best-fitting model with high precision.

\subsubsection{Consideration of Rotation}
\label{ssec:rotation}
Since HD\,139614 is an A-type star and its equatorial rotation rate is not known, it is prudent to consider rotation as an input parameter in the models where possible.
Before ultimately restricting to the pre-main sequence, our coarse grids included variations in surface rotation velocity: $v_{\rm surf} = [0,20,80, 160, 250]$\,km\,s$^{-1}$. Ignoring age priors and $\Delta \nu$ entirely, the preferred surface rotation rates (according to a weighting scheme involving only agreement with temperature and luminosity) were found to be in the range of 60--80\,km\,s$^{-1}$, consistent with calculations based on $v \sin i$ and disc inclinations in Sec.\,\ref{sec:lit}. 

To calculate the impact of oblateness, we computed the rotating models' polar, equatorial, and average values for effective temperature and luminosity. The difference between the polar and equatorial temperature is 30\,K for a MESA model rotating at $v_{\rm eq} = 80$\,km\,s$^{-1}$ that has the same \teff\ and \logL{} as HD\,139614. For the luminosity, the difference in the log is 0.012. These values are small compared to the observational uncertainties. Ultimately, we explored a wide parameter space surrounding our best seismic fit (see discussion in Sec.\,\ref{ssec:global_fits}), so oblateness had no effect on our search for viable models. We discuss its (negligible) impact on the results in Sec.\,\ref{sec:results}. This is despite the fact that we were unable to include rotation in our models directly. Pre-built stellar models with rotation are already evolved to an age older than the literature values for HD\,139614, and as such, the best-fitting rotating models all had ages $>$ 0.8\,Gyr and asymptotic values of $\Delta\nu$ were never within 20\% of their expected value. We therefore had to calculate the pre-main sequence evolution directly, but calculations that included rotation never converged to hydrostatic models. As such, the pre-main sequence grid ultimately did not include rotation.

\subsection{Seismic Models}
\label{ssec:seismic_models}

MESA estimates $\Delta\nu$ for each model by integrating the sound speed across the star. According to the asymptotic theory of stellar oscillations, this value approaches the actual spacing of p\:modes in the limit of high radial orders.
However, the modes in \dsct\ stars have $n=1$ to approximately 8, and so it is preferable to solve the oscillation equations directly, which provides accurate frequencies at all $n$.  We performed these calculations with GYRE \citep{townsend&teitler2013,townsendetal2018}.

GYRE is integrated with MESA but independently maintained and can be used as a stand-alone package. 
It uses the Magnus Multiple Shooting scheme to solve the oscillation equations \citep{kiehl1994, gander&vandewalle2007}. 
We performed a scan using a linear grid with \texttt{alpha\_osc = 10} and \texttt{alpha\_exp = 2} \citep{GYREdoc} over the frequency range $0.5 - 75$\,d$^{-1}$. Our calculations were adiabatic, we considered pressure and gravity modes with any $n$ for $l = 0,1,2,3$, and $m$ was taken to be zero for all modes.

We computed a set of synthetic frequency spectra with GYRE for every model in the pre-main sequence grid that was consistent with the classical constraints.
In total, over 13\,300
synthetic frequency spectra were computed out of nearly 4 million total evolutionary points in the set of stellar tracks. We compared the calculated frequencies to F1 through F11 under ten scenarios for the mode identifications. The identifications given in Table\:\ref{tab:puls} gave the best fit to the models (see Sec \ref{ssec:mode_id} for details). Following \citet{beddingetal2020}, we did not correct the model frequencies for near-surface effects in the way that is usually done for solar-like oscillations \citep{kjeldsenetal2008,ball&gizon2014}. It seems that the models fit without such corrections, probably because A stars have thin surface convection zones. We return to the discussion of atmospheric boundary conditions in Sec.\,\ref{ssec:global_fits}.

\subsection{Determining Global Best Fits}
\label{ssec:global_fits}
\begin{figure}
\begin{center}
\includegraphics[width=0.48\textwidth]{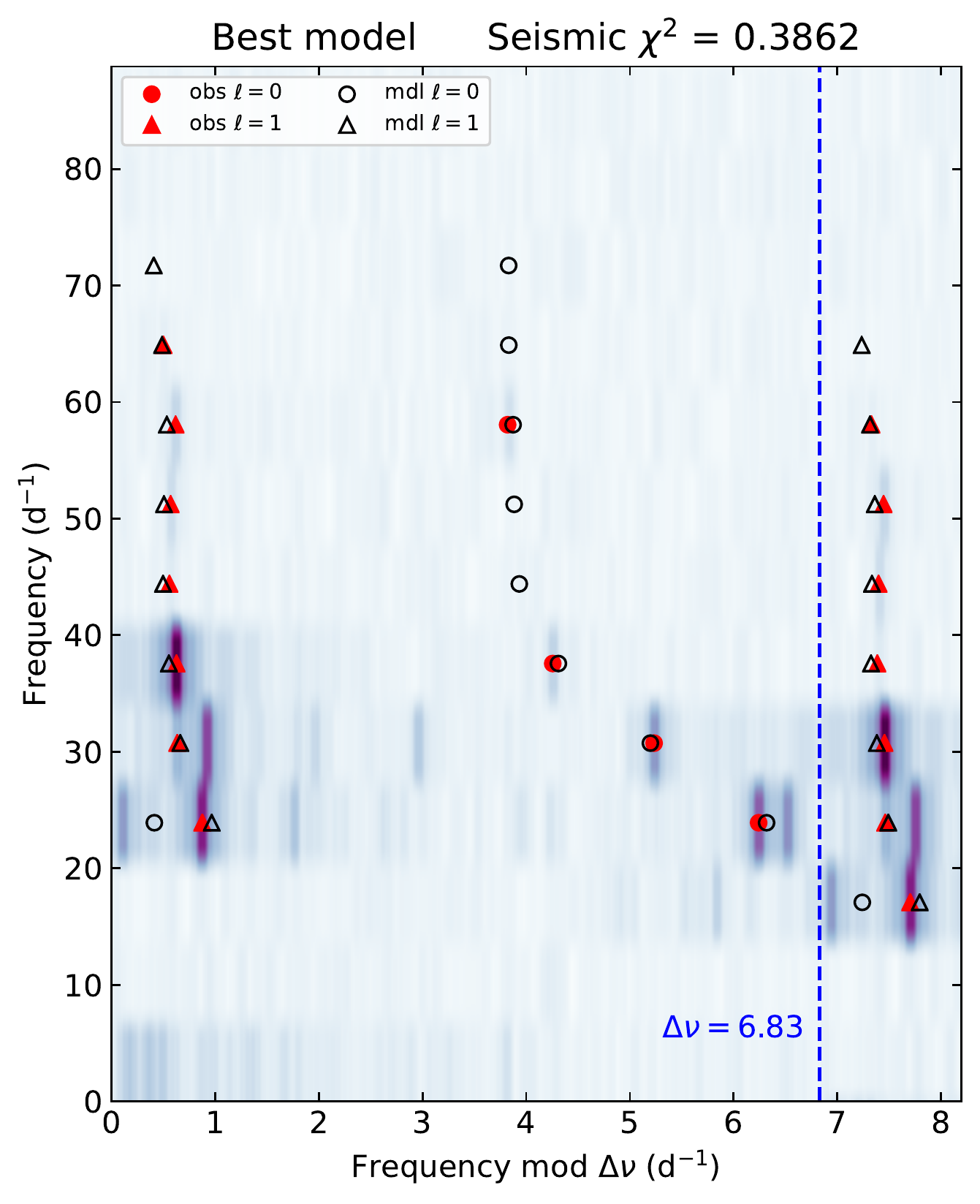}
\caption{\'Echelle diagram showing the best asteroseismic model for HD\,139614, folded with the empirically derived $\Delta\nu$ of 6.83\,d$^{-1}$ found here and in \citet{murphyetal2020b}.}
\label{fig:echelle_model}
\end{center}
\end{figure}

The quality of seismic fits was assessed by generating \'echelle diagrams, with each theoretical frequency spectrum superimposed on the observations.  Our statistically preferred model is shown in Fig.\,\ref{fig:echelle_model}, and we can see it has excellent agreement.
Formal, quantitative agreement was assessed using a pseudo-$\chi^2$ cost statistic.

We assigned uniform uncertainties of $0.1$\,d$^{-1}$ to each frequency because (i) the reported observational uncertainties for the individual frequencies are very small; (ii) uncertainty in mode ID is taken into account via the consideration of multiple ID assignment scenarios; and (iii) there is no reason to prioritize agreement with any particular mode over any other (of the core 11 frequencies considered). This avoids the fit being dominated by the strongest few modes. This uncertainty of 0.1\,d$^{-1}$ corresponds roughly to that of the weakest identified modes and to the symbol sizes in our \'echelle diagrams.

For each seismic model, we computed the normalized, pseudo-$\chi^2$ agreement statistic:\footnote{This is not a true $\chi^2$ because the underlying distribution is not strictly normal, nor are the modeled frequencies independent.}
\begin{eqnarray}
\label{eq:chisq}
\chi^2 = \frac{1}{N}\sum_{i=1}^N \left( \left[ \frac{F_{t,i} - F_{o,i}}{\sigma_{F,o_i}} \right]^2  \right).
\end{eqnarray}
In this statistic, which we refer to as the ``seismic $\chi^2$,'' $F$ are the individual frequency values, $N$ is the number of frequencies considered, and $o$ and $t$ subscripts refer to observed and theoretical values, respectively. The seismic $\chi^2$ for our models ranged from roughly 0.3 to over 50\,000. Comparison with \'echelle diagrams confirmed that the statistically preferred models also gave the best visual fits. 

%

\begin{figure}
\begin{center}
\includegraphics[width=0.48\textwidth]{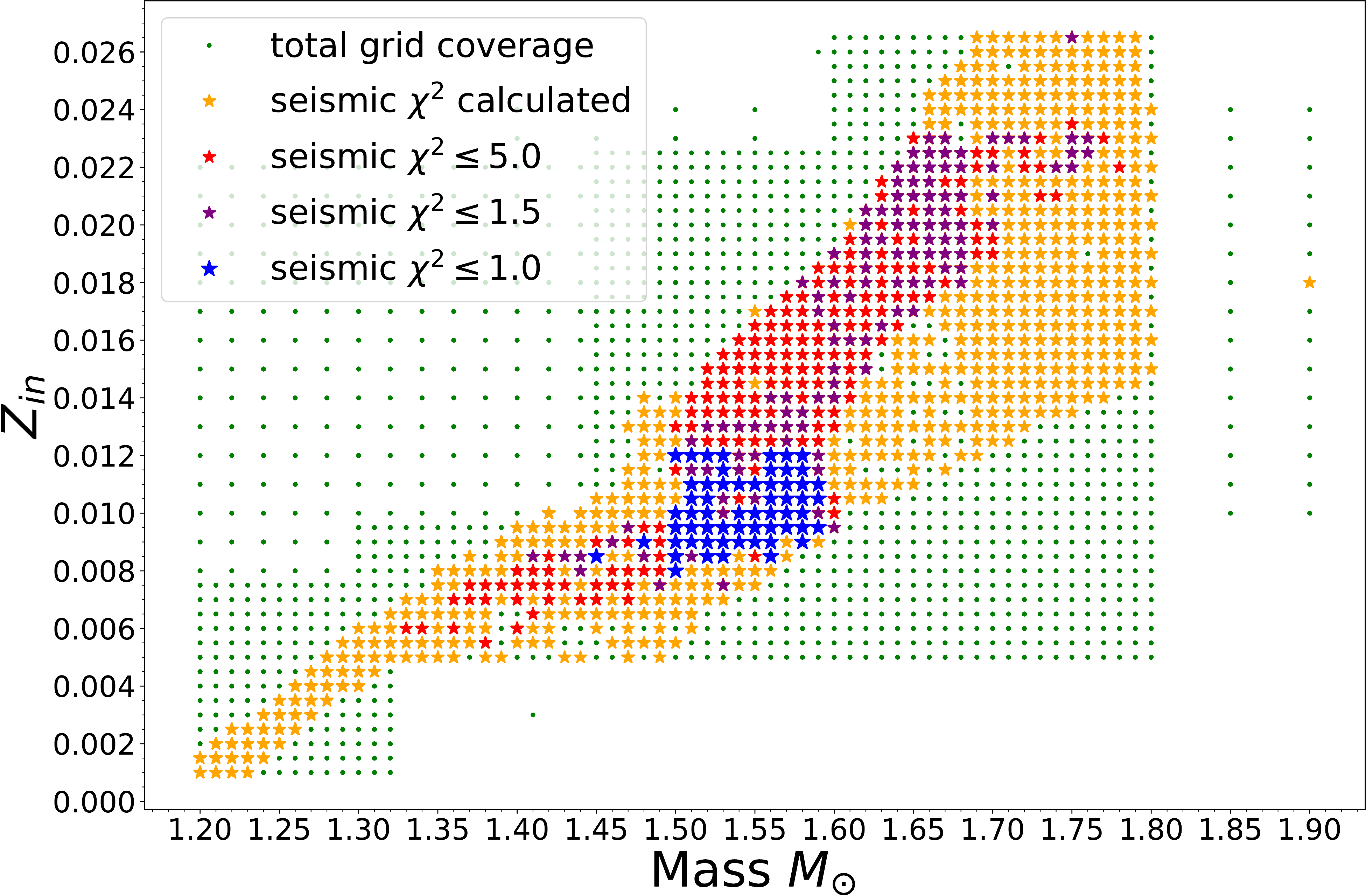}
\caption{The grid of evolutionary tracks considered is shown in terms of mass and metallicity. Each point may represent multiple models with different parameters (e.g. ages), in which case the lowest seismic $\chi^2$ is shown. Unequal grid spacing is due to the use of an adaptive search method \citep{joyce&chaboyer2018a}.}
\label{fig:gridmap}
\end{center}
\end{figure}

Models were initially calculated in broad a grid surrounding the classical constraints and that grid was expanded adaptively to find the global minimum seismic $\chi^2$ by reference to heatmaps (Fig.\,\ref{fig:gridmap}), ensuring the best solution was not bounded by the edge of the computational grid. Consistency with the classical parameters was then re-evaluated when inferring the best-fitting stellar parameters (Sec.\,\ref{sec:results}).

Once the global minimum in seismic $\chi^2$ was found, we tested the effects of different stellar atmospheres on the $\chi^2$ value, calculating new evolutionary tracks with the same mass, mixing length, metallicity, and helium abundance. The age was resampled using the new tracks. All of {\tt simple\_photosphere}, {\tt Krishna\_Swamy}, {\tt photosphere\_tables} and {\tt grey\_and\_kap} atmospheres gave similar $\chi^2$ minima at ages within 0.25\,Myr of each other (i.e. to well within the 1$\sigma$ uncertainties we derive in Sec.\,\ref{sec:results}). Similar conclusions have been reached for other classes of pulsators even more sensitive to atmosphere boundary conditions \citep{yildiz2007,joyce&chaboyer2018b,nsambaetal2018,vianietal2018}. We tested different nuclear nets similarly. For a 1.5-M$_{\odot}$ star, most deuterium burning occurs in the first 1\,Myr of the stellar evolution and by 10\,Myr the effects of this burning on the evolution are negligible \citep{krumholz2011,krumholz2014}. We recalculated evolutionary and seismic models for our best-fitting model, adding the {\tt pp\_extras} nuclear net and adopting a deuterium-to-hydrogen mass fraction of $2\times10^{-5}$ \citep{stahleretal1980,linsky1998}. This nuclear net more accurately represents nuclear reactions in young stars but the resulting change in the stellar parameters of HD\,139614 was insignificant. The $\chi^2$ minimum was of comparable depth at an age of 10.73\,Myr, i.e. adding deuterium burning resulted in an age difference of only 0.02\,Myr, much smaller than our quoted 0.77\,Myr uncertainty.


\section{Results}
\label{sec:results}

\begin{figure*}
    \centering
    \includegraphics[width=\textwidth]{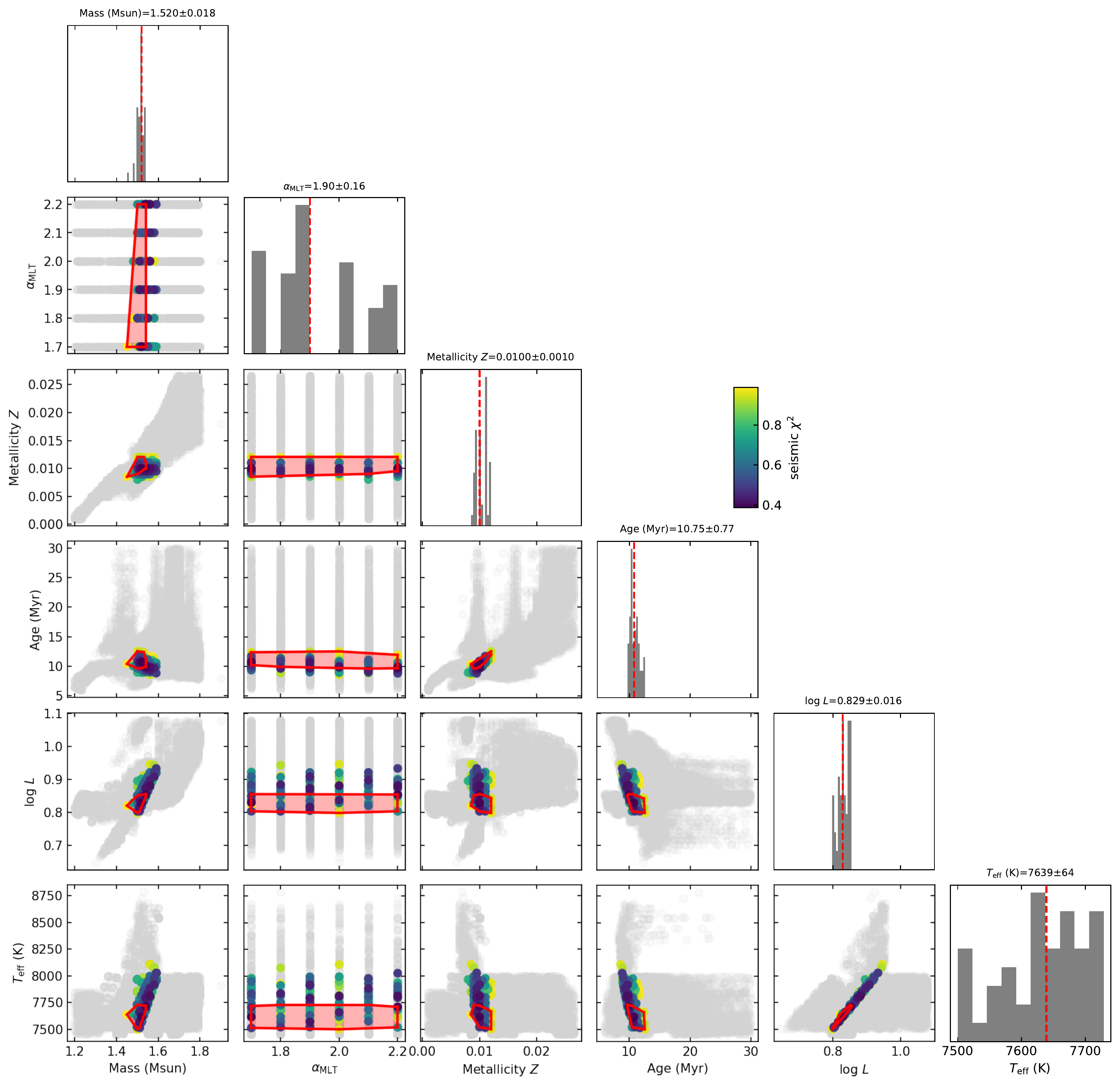}
    \caption{Corner plot of the modelling results. Grey circles show all models for which a seismic $\chi^2$ was calculated, and coloured circles show the 125 models that have seismic $\chi^2<1$, with darker colours representing lower values. The shaded red area encloses the subset of those 125 models that lie within the $1\sigma$ error box in $T_{\rm eff}$ and $\log L$. Histograms are calculated on that 47-model subset, the median and standard deviation of which are written.}
    \label{fig:corner}
\end{figure*}

We evaluated the 13\,000 models with calculated pulsation spectra to find the global minimum in our seismic $\chi^2$ statistic and thereby find the best-fitting stellar parameters. Note that the seismic $\chi^2$ is not normalised (it is set by the adopted uncertainties in frequencies used for modelling). We found that the global minimum could be isolated by setting a threshold of $\chi^2<1$, which we adopted in our analysis. A broad overview of the investigated models and the dependence of the seismic $\chi^2$ on the various stellar parameters is shown in Fig.\,\ref{fig:corner}. The best models are well localised for all parameters except $\alpha_{\rm MLT}$, from which we conclude that the results do not depend on the mixing length.

\begin{table}
    \centering
    \caption{Modelling results for HD\,139614. The best model (lowest seismic $\chi^2$ statistic, 0.3862) also coincides with the medians of the parameter distributions, quoted here. Uncertainties are the standard deviations of samples with seismic $\chi^2 \leq 1.0$. $^{\dagger}$This asymptotic value of $\Delta\nu$ exceeds the empirical value for reasons explained in Sec.\,\ref{ssec:seismic_models}.}
    \begin{tabular}{lcrl}
    \toprule
    Parameter & Units & Value & Uncertainty \\
    \midrule
    mass & M$_{\sun}$ & 1.520 & 0.018 \\
    $Z$ & & 0.0100 & 0.0010 \\
    age & Myr & 10.75 & 0.77 \\
    $\log L/{\rm L}_{\sun}$ (avg) & & 0.829 & 0.016 \\
    $T_{\rm eff}$ (avg)\tablefootnote{We quote the precision determined from modelling, only. Absolute accuracy is limited to 2\% by the calibration of the $T_{\rm eff}$ scale by interferometry \citep{casagrandeetal2014,whiteetal2018}} & K & 7615 & 64 \\
    $\alpha_{\rm MLT}$\tablefootnote{No dependence found. The best-fitting model happened to have \mbox{$\alpha_{\rm MLT}=1.9$.}} & & 1.9 & --- \\
    $\Omega$ & km\,s$^{-1}$ & 0.0000 & (fixed) \\
    $Y$ & & 0.29 & (fixed) \\
    $\Delta\nu$ (MESA, asymptotic$^{\dagger}$) & $\upmu$Hz & 87.0670 & \\
    $\Delta\nu$ (MESA, asymptotic$^{\dagger}$) & d$^{-1}$ & 7.523 & \\
    \bottomrule
    \end{tabular}
    \label{tab:results}
\end{table}

We determined the best-fitting values by calculating the median of each parameter distribution, using the models inside the classical error box with seismic $\chi^2<1$. We quote 1$\sigma$ uncertainties from the standard deviation of those models. With the exception of mode IDs, we used only a single set of fixed modelling assumptions, so the uncertainties do not take into account modelling systematics or theoretical uncertainties in the physical prescriptions themselves. The results are given in Table\:\ref{tab:results}. The best fitting model, with $\chi^2 = 0.3862$, happens to have $Z$, age, and mass equal to those median values. This is the model shown in Fig.\,\ref{fig:echelle_model}.

In the following sections we describe key results for individual parameters, starting with metallicity, and discuss our $\chi^2<1$ threshold.

\subsection{Metallicity}

\begin{figure}
    \centering
    \includegraphics[width=\columnwidth]{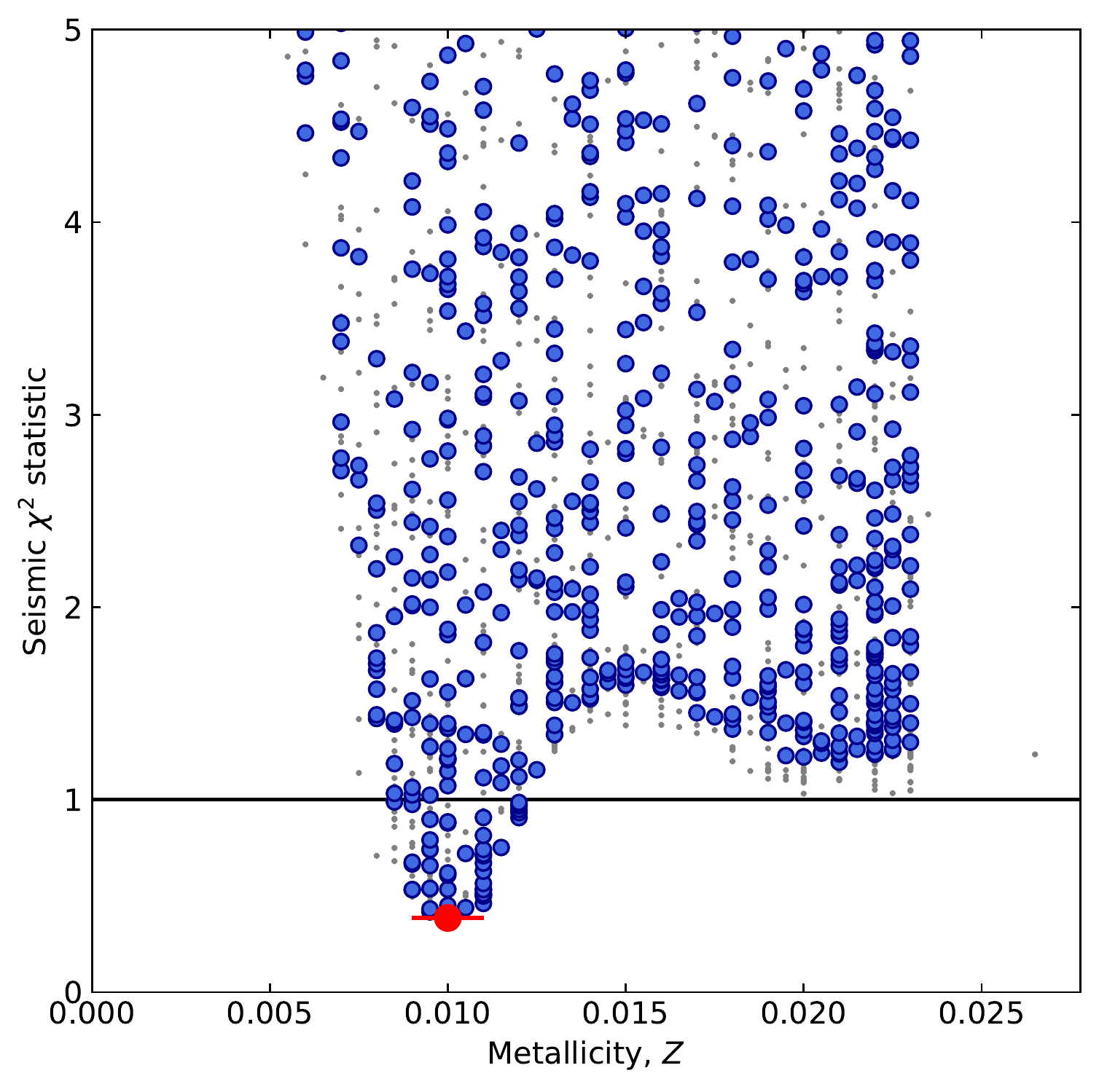}
    \caption{The seismic $\chi^2$ of the best models as a function of the metallicity, $Z$. Models within the classical box are shown in blue, whereas models outside this box are smaller and grey. The black line is our threshold at $\chi^2=1$. The red circle and error bar show the median and 1$\sigma$ values determined.}
    \label{fig:z}
\end{figure}

Fig.\,\ref{fig:z} shows the seismic $\chi^2$ values as a function of metallicity, which have a well-defined minimum at $Z=0.0100\pm0.0010$. Comparison with Fig.\,\ref{fig:corner} shows that the right-hand edge in the metallicity distribution in Fig.\,\ref{fig:z} is a real drop-off in seismic agreement and does not represent an edge of the computational grid. Similarly, metal weak models having $Z\lesssim0.007$ are strongly ruled out by seismology alone. 
We have made the first asteroseismic determination of the interior metallicity of a \LB\ star, and we found it is not metal weak.

\subsubsection{Present-day surface abundance vs global composition}



Since the chemical composition of any star is stratified and varies with time, the $Z$ we have determined is not necessarily representative of the observed $Z$, even in the absence of chemical peculiarities. Choices for model parameters governing mixing processes, especially diffusion, will therefore affect the evolution of the surface abundance \citep{thoul1994,dotteretal2017}. Given the young age of HD\,139614, however, the timescale over which diffusion acts is small. Nonetheless, we emphasize that our models do not include diffusion and that the metallicities discussed in our results represent the initial, global metal abundance and should not be conflated with a spectroscopic surface abundance.

\subsection{Age}

\begin{figure}
    \centering
    \includegraphics[width=\columnwidth]{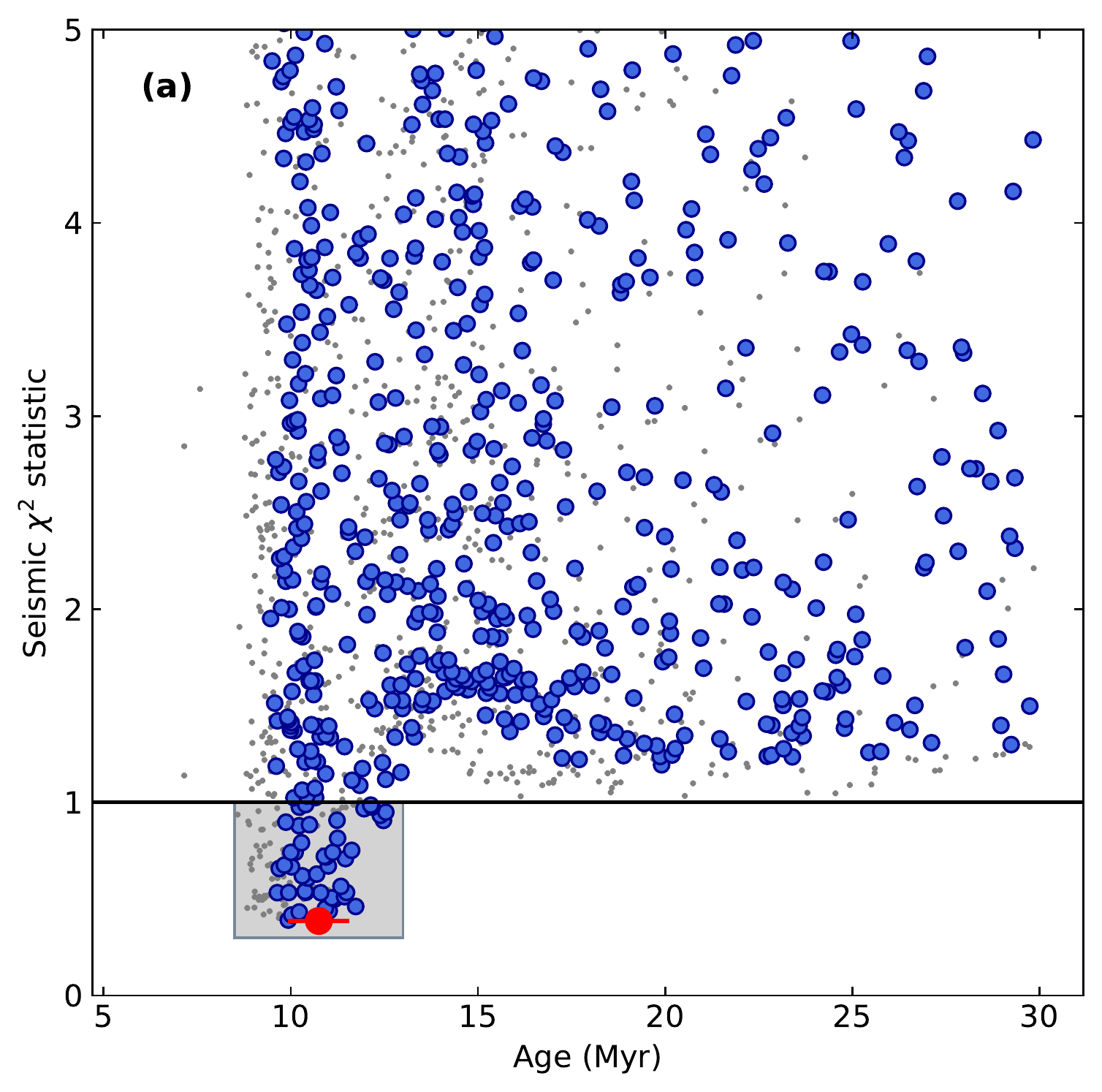}
    \includegraphics[width=\columnwidth]{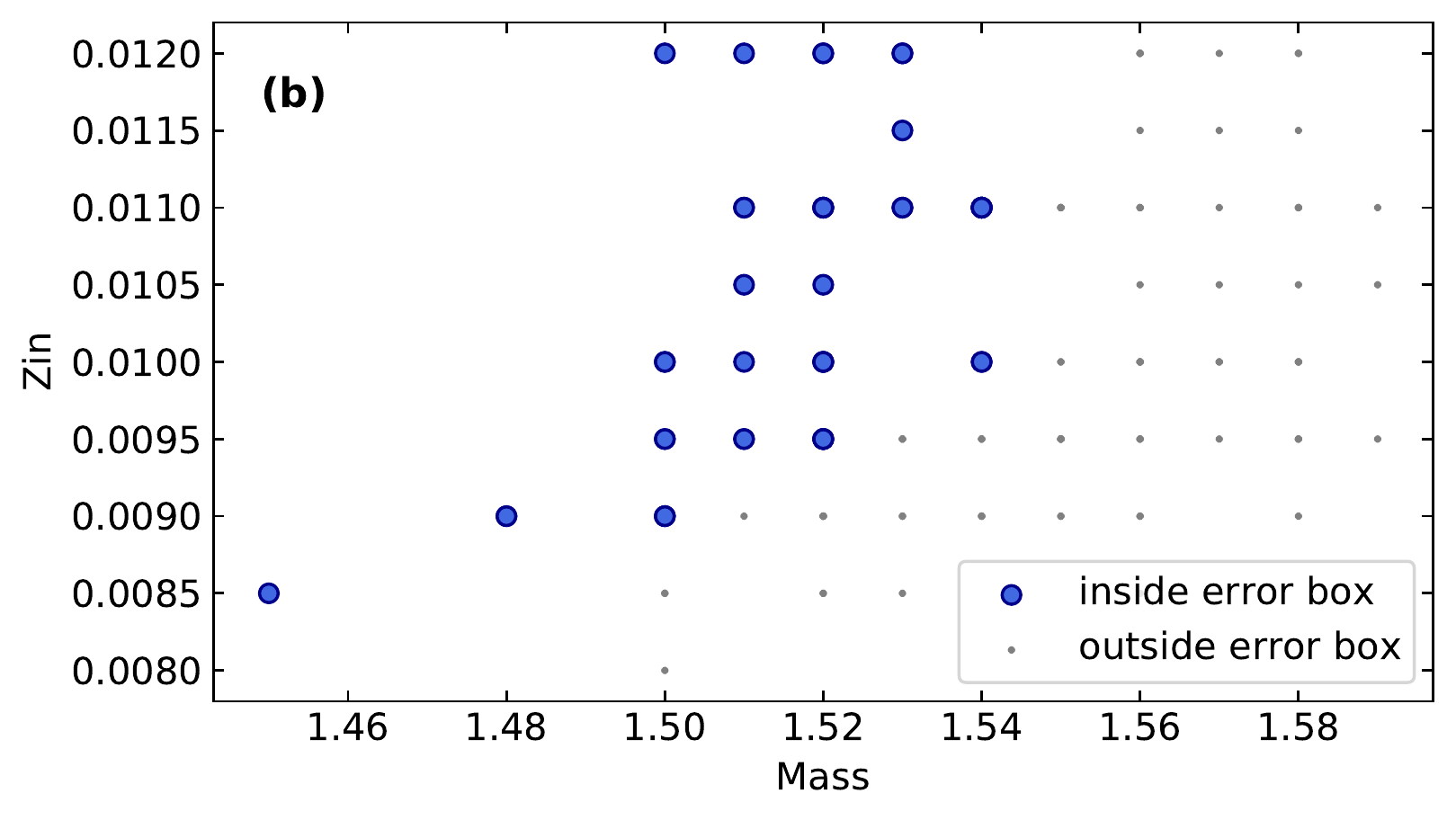}
    \caption{{\bf (a):} The seismic $\chi^2$ of the top models as a function of the stellar age. Models within the classical box are shown in blue, whereas models outside this box are smaller and grey. The black line is our threshold at $\chi^2=1$. The red circle and error bar show the median and 1$\sigma$ values determined. {\bf (b):} The mass and metallicity of those models in the shaded region of panel a. More than one model can occupy each point in this plane.}
    \label{fig:age}
\end{figure}

Fig.\,\ref{fig:age}a shows the seismic $\chi^2$ values as a function of age, which also have a well-defined minimum ($10.75\pm0.77$\,Myr) separated by the threshold of $\chi^2$ = 1. However, there is a division between models inside and outside the classical box among the models with $\chi^2<1$, which occurs at 9.5\,Myr. Fig.\,\ref{fig:age}b shows that this division corresponds to an edge in mass and metallicity: the more massive and lower metallicity models fall outside the classical box. This division also extends to models of higher $\chi^2$, and is visible in Fig.\,\ref{fig:corner}.

Fig.\,\ref{fig:age}a also shows a sharp edge at 8.8\,Myr where seismic agreement rapidly falls off. Fig.\,\ref{fig:corner} shows that this is not an edge effect in our computational grid: models with younger ages were calculated (down to $\sim$5\,Myr), but these do not provide good seismic matches. Had the models between 8.8 and 9.5\,Myr not been excluded by the classical error box, the inferred age would differ by less than $1\sigma$ because of the strong seismic constraints.

\subsection{Mass}

\begin{figure}
    \centering
    \includegraphics[width=\columnwidth]{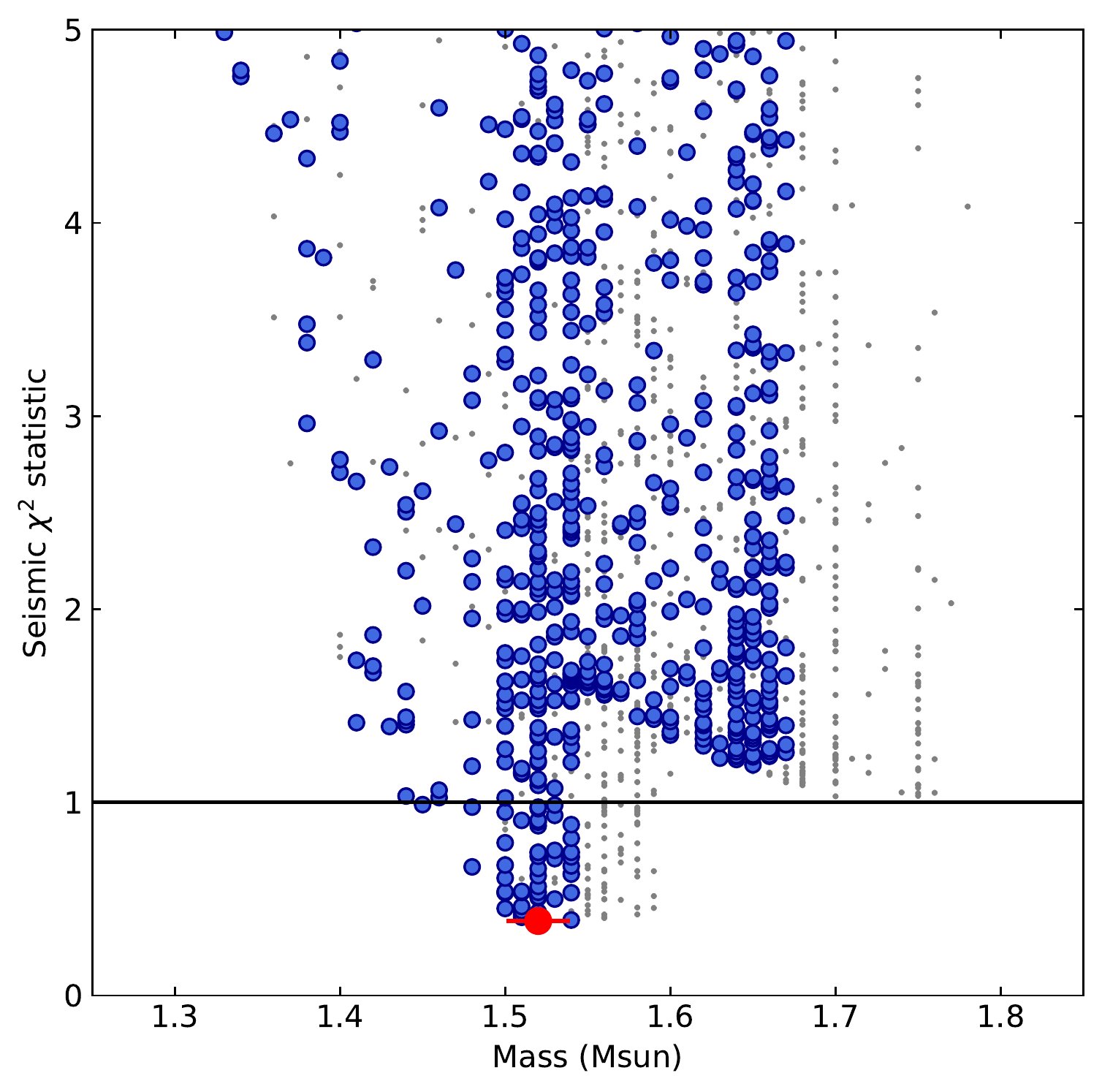}
    \caption{The seismic $\chi^2$ of the top models as a function of the stellar mass. Models within the classical box are shown in blue, whereas models outside this box are smaller and grey. The black line is our threshold at $\chi^2=1$. The red circle and error bar show the median and 1$\sigma$ values determined.}
    \label{fig:mass}
\end{figure}

The mass range of the global minimum in seismic $\chi^2$ occurs at 1.5--1.6\,M$_{\odot}$ (Fig.\,\ref{fig:mass}), but the low-$\chi^2$ models above 1.55\,M$_{\odot}$ are not consistent with classical constraints. Using all models inside the classical box with seismic $\chi^2<1$, we determine the stellar mass as $1.520\pm0.018$\,M$_{\odot}$. We checked the impact of neglecting oblateness, which equates to a maximum change of 0.012 in \logL{} (Sec.\,\ref{ssec:rotation}): extending the classical box by this amount does not cause any additional models to fall inside it.

\section{Discussion}
\label{sec:discussion}

\subsection{Mode identification ambiguity}
\label{ssec:mode_id}

HD\,139614 is on the pre-MS, where pulsation properties change rapidly (Fig.\,\ref{fig:dnu_age}). Our mode identification is based on lessons learned in \citet{beddingetal2020}, where stars were all thought to be on the MS. We must therefore be careful in extrapolating to the pre-MS regime, and look carefully at our assumptions.

Our final mode identification used GYRE outputs to refine an initial mode identification based on ridges in the echelle diagram. While these are clear and well-separated at high frequencies, there is some ambiguity at lower frequencies, in the orders with \mbox{$n=1$ or 2}. We initially assumed that the radial and dipole modes were the strongest peaks lying near to each ridge. This has physical motivation: mode visibility on an unresolved stellar disc diminishes rapidly from $\ell\geq2$ \citep{aertsetal2010}, so one does not generally expect $\ell\geq2$ modes to have higher amplitudes than $\ell \leq 1$ modes. However, the latter are not necessarily {\it driven} as strongly as the former, which also affects the observed amplitudes.

An example of a lower-amplitude peak being a better fit to the ridge is the $n=2$ dipole mode. If the stronger peak here is adopted instead (F12 instead of F2), the dipole ridge is made less straight, in conflict with the models. The $n=2$ radial mode is also slightly ambiguous. We considered that the peak lying at slightly higher frequency (F13) could be the radial mode instead of F8, but this caused disagreement in the frequency ratio of the $n=2$ and $n=3$ radial modes, as well as the frequency difference with the $n=2$ dipole mode.

At $n=1$, there are two strong peaks that we might identify: F1 and F14. F1 lies at the base of the dipole ridge, hence we identified it as the $n=1$ dipole mode. We considered the possibility that this could be the fundamental radial mode instead (in accordance with the identification in \citealt{murphyetal2020b}), but this was a much worse fit in all models. Ultimately, the fundamental radial mode is not assigned to any mode, as it does not lie close to F14 either. This could be considered a weakness of our analysis, but it is also possible that the fundamental mode is simply not excited in this star. None of the seven peaks (F14--F20) at frequencies lower than the $n=1$ dipole mode were identified, including among g-mode frequencies we calculated. Some of these unidentified modes could be f modes (e.g. \citealt{cowling1941,stelloetal2014}) or island modes (e.g. \citealt{lignieres&georgeot2009,bouchardetal2020}). Other observed peaks that remain unidentified (F21--F35) could belong to modes of higher degree, but there are many more peaks at radial orders of $n=1$ and 2 than can be explained simply by including $\ell=2$ and $\ell=3$ modes. They could also be modes of different azimuthal order that our non-rotating models are unable to identify.

In our modelling, we have placed importance on some high-frequency modes of low SNR. Pulsation models lack the ability to predict observed amplitudes, but the mode frequencies remain relevant for constraining the internal structure of the star, regardless of the amplitude to which they are driven. The fact that these low SNR modes lie on ridges in the \'echelle and are reproduced by the models is evidence that the identifications are correct. One of our mode ID scenarios used only the higher-frequency modes that have the same ID in every scenario (i.e. with none of the ambiguity mentioned above). The top models ($\chi^2 < 1$) under that scenario had median values of mass, age and metallicity within the 1$\sigma$ values quoted in Table\:\ref{tab:results}.

\subsection{Comparing pre-ZAMS and post-ZAMS mode frequencies}
\label{ssec:pre_vs_post}

For the final 10--20 Myr before the ZAMS, the evolutionary tracks of pre-MS stars lie very close to or cross their post-ZAMS counterparts (Fig.\,\ref{fig:hr_prems}). \citet{suranetal2001} compared the pulsations of 1.8-M$_\odot$ pre-MS and post-ZAMS models, finding that the $\ell=0$ and high-frequency $\ell=1$ and $\ell=2$ modes were indistinguishable near where the tracks cross. This is because the models have similar mean densities and outer layers. In contrast, \citet{suranetal2001} also found that there were significant differences in low-frequency $\ell=1$ and $\ell=2$ modes. These are g\:modes or, in some cases, mixed modes, which are sensitive to the deep internal structure. Observations of these low frequency non-radial modes would therefore allow pre-MS and post-ZAMS stars to be distinguished from seismology alone. We now discuss whether we are able to do this for HD\,139614.

\begin{figure}
    \centering
    \includegraphics[width=0.47\textwidth]{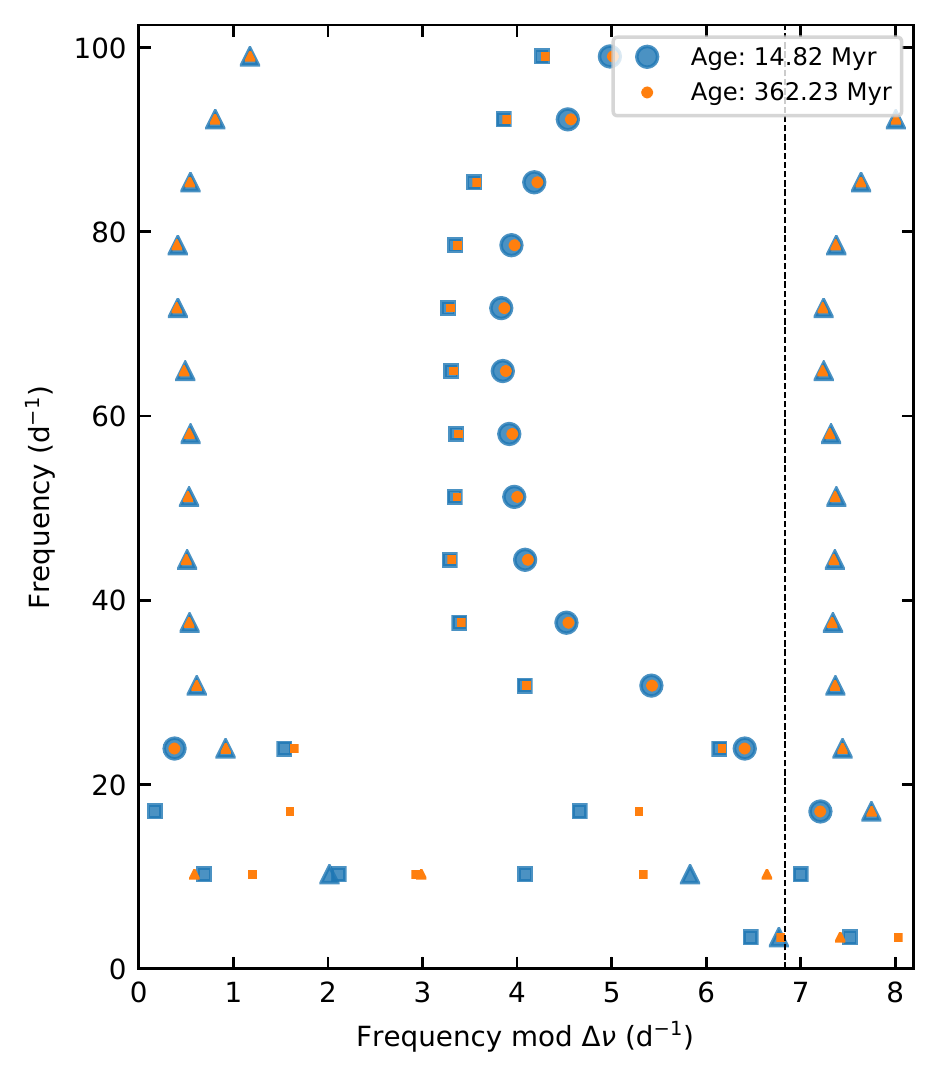}
    \caption{Pre-MS models (blue) can be very similar to post-ZAMS models (orange), shown here for a 1.6\,M$_{\odot}$ star and $\Delta\nu = 6.83$\,d$^{-1}$. Circles, triangles and squares show $\ell=0, 1$ and 2 modes, respectively.}
    \label{fig:pre_vs_post}
\end{figure}

Fig.\,\ref{fig:pre_vs_post} shows an \'echelle diagram for a 1.6-M$_\odot$ model at the pre-MS and post-ZAMS ages where the model has the same \Dnu\ as HD\,139614. As expected, major differences between the models are seen at low frequencies. The modes below the frequency of the radial fundamental mode are g\:modes, which are only sometimes observed alongside $\delta$\,Sct pulsations (e.g. \citealt{zhangetal2020}), and we have not detected them in HD\,139614. These models also have mixed $\ell=2$ modes in the region where unidentified modes have been detected in HD\,139614. However, with the number of unidentified modes exceeding the expected number of $\ell=2$ mixed modes, and no ridge of high-frequency $\ell=2$ p-modes to use as guides, we are unable to reliably identify any of these peaks as $\ell=2$ mixed modes. Finally, while there are also subtle differences in the pulsation frequencies on the $\ell=0$ and $\ell=1$ ridges, these are observationally indistinguishable at this level. Therefore, external age constraints such as our $<$30-Myr age prior from UCL membership are necessary to distinguish between the two evolutionary phases \textit{for HD\,139614}. This then casts doubt on how reliably pre-MS and MS stars can be distinguished by their pulsation spectra alone in cases similar to this where no g\:modes are detected, and no $\ell=2$ modes  are identified.

\begin{figure}
    \centering
    \includegraphics[width=0.47\textwidth]{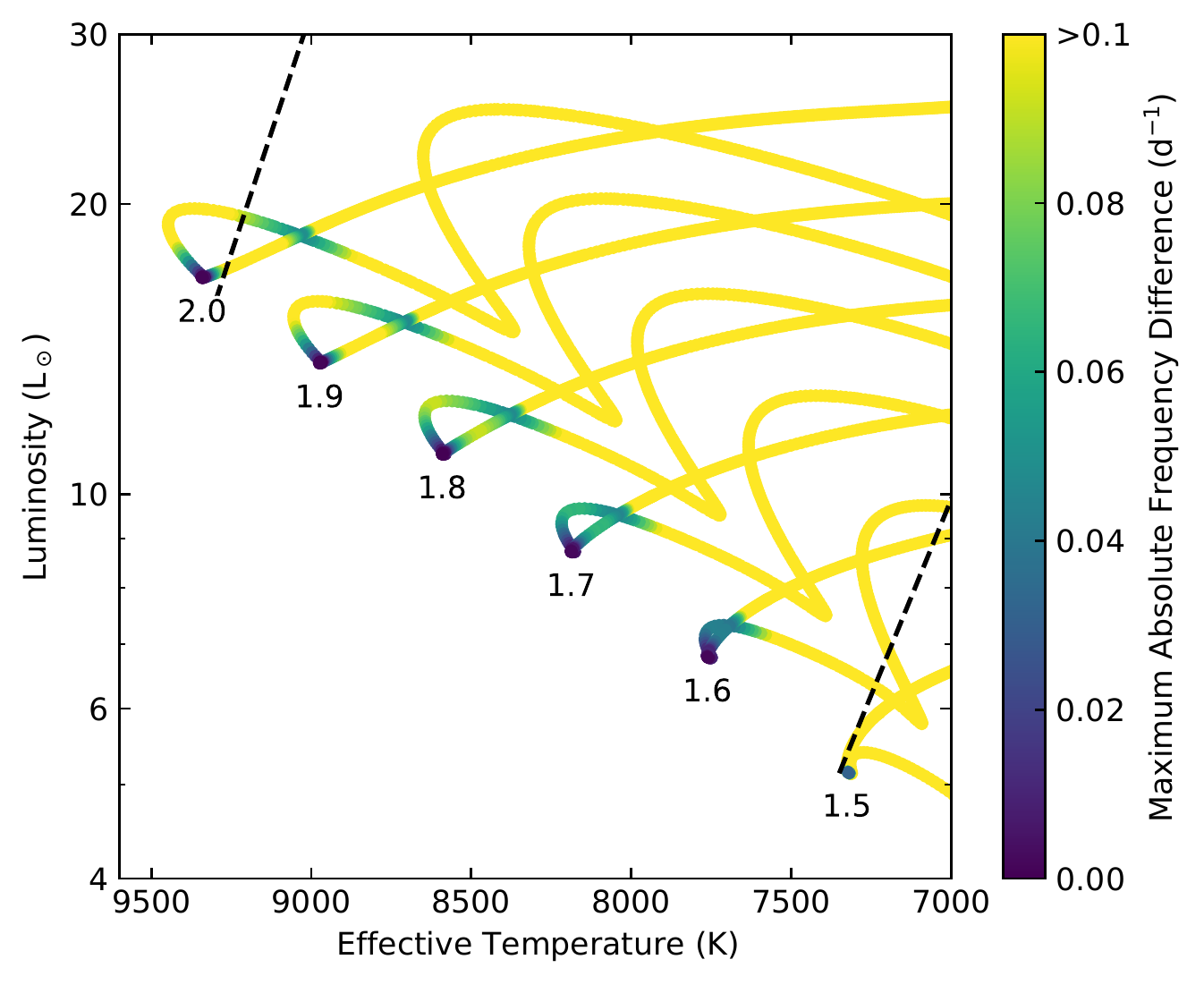}
    \caption{HR diagram indicating regions where it may not be possible to distinguish between pre-MS and post-ZAMS models from seismology alone. The colour scale indicates the largest absolute frequency difference of any \mbox{$\ell = 0$ or 1} p\:mode with $n\le9$ between closest matching pre-MS and post-ZAMS models. The summed difference over many frequencies may be much greater, and significant under a $\chi^2$-like statistic. The dashed lines indicate the empirical boundaries of the $\delta$\,Sct instability strip from \citet{murphyetal2019}.}
    \label{fig:freqdiff_HR}
\end{figure}

Fig.\,\ref{fig:freqdiff_HR} shows the regions in the HR diagram where there is substantial overlap in the $\ell=0$ and $\ell=1$ frequencies of pre-MS and post-ZAMS models. This is confined to a small area where the pre-MS crosses the main sequence just prior to the ZAMS, and near the ZAMS itself. However, depending on the stellar mass, the duration of this near-overlapping phase of the post-ZAMS track can be a few-hundred Myr, up to about 20 per cent of the MS lifetime. If an age prior is available, such as via membership in a stellar association, precise asteroseismic masses, ages and metallicities may be determinable for young stars over a wide range of masses.

Similar crossings of pre-MS and MS tracks occur further from the ZAMS. In these instances the MS star will be more evolved and its $\ell=1$ frequencies will exhibit bumping of mixed modes and other effects from chemical discontinuities near the core. Since these will be absent in the homogeneous pre-MS star, evolutionary phases of stars at these other crossings, which include those modelled by \citet{guentheretal2007}, should therefore be distinguishable.

\subsection{Systematic age uncertainty}
\label{ssec:age_systematics}

The dominant sources of age uncertainty in stellar modelling are, in no particular order, (i) heavy element diffusion, which can deliver fresh nuclear fuel into the star's core; (ii) the size of the convective core, as influenced in models by the degree of convective overshoot (pressure scale heights) and the prescription for convective overshoot (exponential vs linear); (iii) bulk material transport, which can also deliver fresh fuel to the core; and (iv) the mixing length $\alpha_{\rm MLT}$, which determines the efficiency of the energy transport mechanisms throughout the model and the ratio of flux transported by each of them, by changing the size of the convection zone(s). None of these four sources contributes substantial age uncertainty in our model of HD\,139614, primarily because of its young age. The degree of mixing of nuclear fuel into the star's core is irrelevant when the star's main energy source is gravitational contraction and the measurement of its age hinges on the degree of that contraction to the ZAMS, rather than the amount of its nuclear fuel that remains. We have already shown in Sec.\,\ref{sec:results} that $\alpha_{\rm MLT}$ has no impact on our results. On the other hand, episodic accretion is known to contribute a substantial fractional age uncertainty for young low-mass stars \citep{baraffeetal2009,vorobyov&elbakyan2020}, but this is not well quantified for intermediate-mass stars, which are modelled with different atmospheres and opacities. With the exception of atmospheric boundary conditions, we have not investigated systematic modelling uncertainties in this work, and do not include them in our quoted random uncertainties. However, we have considered the inaccuracy between different age=0 definitions, which is of order $10^4$\,yr \citep{soderblom2010} and therefore insignificant relative to our uncertainty of $\sim$$10^6$\,yr (Sec.\,\ref{ssec:tracks}; see also \citealt{zwintz2020,steindl&zwintz2020}).

\subsection{Uncertainty from rotation}
\label{ssec:rot_uncertainty}

A potential source of systematic error in this work is the lack of inclusion of rotation into our models. Rotation breaks the spherical symmetry of the different azimuthal orders of non-radial pulsation, which leads to rotational splittings when the star is observed from moderate inclinations. We see the star nearly pole-on, so no splittings are observed, but asphericity has other effects. One is gravity brightening at the poles, making the observed stellar temperature and luminosity a function of viewing angle. This has already been discussed in Sec.\,\ref{ssec:rotation}, and the magnitude of its effect has been calculated and deemed insignificant to the methodology and results. Another effect of asphericity is to shift pulsation frequencies. For an equatorial velocity of 80\,km\,s$^{-1}$, the frequency shifts are expected to be no more than 0.5\% for radial orders $3\leq n \leq 10$, which equates to little more than what our uncertainties already encompass. For modestly larger inclinations ($30^{\circ}$ instead of 20$^{\circ}$), $v_{\rm eq} = 50$\,km/s$^{-1}$ and the frequency shifts are an order of magnitude smaller \citep{dicriscienzoetal2008}. In summary, we expect that our use of non-rotating models could potentially affect our modelled frequencies by $\sim1\sigma$. 

\subsection{Implications of the determined age of HD\,139614} 

Our determined age of $10.75\pm0.77$\,Myr for HD\,139614 is considerably younger than literature estimates of the age of UCL ($16\pm2$\,Myr, \citealt{pecaut&mamajek2016}; 20\,Myr, \citealt{luhman&esplin2020}). Understanding the origin of this difference is important for understanding the star formation history of the association, the evolution of stars within UCL, and their interactions with their discs.

UCL is so expansive that the assumption of a single age for the entire association is overly simplistic. \citet{pecaut&mamajek2016} showed that there are clear age gradients in each subgroup of Sco-Cen, which for UCL amounts to an age range spanning 10--26\,Myr. In addition to this, any star-forming region can be expected to have an intrinsic age spread of $\sim$2--5\,Myr \citep{soderblom2010,darioetal2016,beccarietal2017,krumholtzetal2019}, and episodic accretion can also cause apparent age spreading \citep{baraffeetal2009,baraffeetal2012}. After accounting for probable age spreads, our age measurement is consistent with association membership, given the specific location of HD\,139614 within UCL. The age precision achieved in this work makes HD\,139614 a useful benchmark within the association.

HD\,139614 is rather old for an object with a protoplanetary disc \citep{ribasetal2015}. Our precise age and bulk metallicity will be useful in understanding the retention and eventual dissipation time of a massive dust disc, which likely goes hand-in-hand with dust trapping \citep{pinillaetal2018}.

\subsection{Future work}

Our model and seismology is a starting point for a more detailed analysis that includes rotation and specific prescriptions of diffusion. That might include a two-zone model with an outer zone with recent accretion, and an inner zone comprising the majority of the stellar mass whose primordial composition has not been modified by selective accretion or mixing. Such an analysis would become even more valuable if a similar but chemically normal $\delta$\,Sct star in UCL could be used for a differential analysis.

Similar analyses could be applied to other stars in UCL in order to construct an asteroseismic age map across the association, or to stars in other associations to fine-tune their ages. \citet{beddingetal2020} have already used this approach to resolve the age debate surrounding the recently-discovered Pisces-Eridanus stream \citep{curtisetal2019,meingastetal2019}. In particular, the ages of many associations are established by the lithium depletion boundary, with which quantitative ages can be difficult to ascertain \citep{soderblom2010}. Asteroseismology would be very welcome to check these Li depletion ages (Zuckerman, priv. comm.).

\section{Conclusions}
\label{sec:conclusions}

HD\,139614 is in a particular evolutionary stage where the asteroseismic large spacing, \Dnu, has good sensitivity to mass. We used MESA and GYRE to leverage this sensitivity and determine a precise asteroseismic age ($10.75\pm0.77$\,Myr) and metallicity ($Z=0.0100\pm0.0010$) without degeneracy with mass ($M=1.52\pm0.02$). We demonstrated that similar analyses may be possible for young stars over a wide range of masses, especially if external constraints on age are available to distinguish whether a given young star lies just before or after the ZAMS (Sec.\,\ref{ssec:pre_vs_post}). Environments such as young stellar associations are therefore particularly well suited to further analyses of this kind, raising the exciting prospect of asteroseismic age benchmarking at the 10\% level for a variety of stellar associations and clusters to help calibrate other age determination methods.

We have made the first asteroseismic determination of the interior metallicity of a \LB\ star, and we found it is not metal weak. The precise bulk stellar metallicity and age determined in this work provide an opportunity to model gas--dust separation and planet formation more precisely, unencumbered by large uncertainties imposed by chemical peculiarities befouling the stellar spectroscopy and by age dispersion and episodic accretion affecting cluster ages. Our snapshot of the evolution of HD\,139614 and its planet-forming disc may also hold clues as to how some stars are able to retain discs for a mysteriously long time, perhaps aided by embedded planets.

Association ages are often determined by main-sequence turn-off of high-mass stars, and/or by study of the lower-mass members, usually revealing mass dependence in the age measurements. The availability of a calibrator at intermediate mass will be helpful to this problem. Similarly, understanding the formation, evolution, and occurrence rates of planets as a function of stellar mass is an active area of research that will benefit from having more precise masses for intermediate-mass stars, where the planet occurrence rate drops precipitously.

\section*{Acknowledgements}
The authors thank Lorenzo Spina, Przemes\l{}aw Walczak, Adam Jermyn, Mark Krumholz, and Jonathan Gagn\'e for discussions. This research made use of {\sc Lightkurve}, a Python package for \textit{Kepler} and TESS data analysis \citep{lightkurvecollaboration2018}, and {\sc echelle} \citep{hey&ball2020}. SJM was supported by the Australian Research Council through DECRA DE180101104. MJ was supported the Research School of Astronomy and Astrophysics at the Australian National University and funding from Australian Research Council grant No.\ DP150100250.
MJ wishes to thank the MESA developers' team and especially Richard Townsend for advice and maintenance of the GYRE forums. MK gratefully acknowledges funding by the University of Tartu ASTRA project 2014-2020.4.01.16-0029 KOMEET, financed by the EU European Regional Development Fund.

\section*{Data Availability}
TESS light curves are publicly available at MAST\footnote{\url{https://mast.stsci.edu/}}. The inlists corresponding to our best-fitting model ({\sc mesa} and {\sc gyre}) are available at the {\sc mesa} hub on Zenodo: \url{https://doi.org/10.5281/zenodo.4291277}.



\bibliographystyle{mnras}
\interlinepenalty=10000
\bibliography{sjm_bibliography} 
\bsp	



\label{lastpage}
\end{document}